\documentclass{natureprintstyle}

\usepackage{graphicx,amsmath} 
\usepackage{amssymb}
\usepackage{color}
 \usepackage{footmisc}

\usepackage{astjnlabbrev-nature}

\newcommand{\appropto}{\mathrel{\vcenter{
  \offinterlineskip\halign{\hfil$##$\cr
    \propto\cr\noalign{\kern2pt}\sim\cr\noalign{\kern-2pt}}}}}

\newcommand{ \fwhm  }{\ifmmode {\text{FWHM} } \else \text{FWHM}\fi} 
\newcommand{ \fwobs  }{\ifmmode {\text{FWHM}_{\rm obs} } \else $\text{FWHM}_{\rm obs}$\fi} 
\newcommand{ \fwobstwo  }{\ifmmode {\text{FWHM}_{\rm obs}^2 } \else $\text{FWHM}_{\rm obs}^2$\fi} 
\newcommand{ \fwobsminus  }{\ifmmode {\text{FWHM}_{\rm obs}^{-1} } \else $\text{FWHM}_{\rm obs}^{-1}$\fi} 

\newcommand{ \fwobshalf  }{\ifmmode {\text{FWHM}_{\rm obs}^{1/2} } \else $\text{FWHM}_{\rm obs}^{1/2}$\fi} 

\newcommand{ \fwint  }{\ifmmode {\text{FWHM}_{\rm int} } \else $\text{FWHM}_{\rm int}$\fi} 
\newcommand{ \fwinttwo  }{\ifmmode {\text{FWHM}_{\rm int}^2 } \else $\text{FWHM}_{\rm int}^2$\fi} 
\newcommand{ \fwintonef  }{\ifmmode {\text{FWHM}_{\rm int}^{1.5} } \else $\text{FWHM}_{\rm int}^{1.5}$\fi}
\newcommand{ \fwintonet  }{\ifmmode {\text{FWHM}_{\rm int}^{1.2} } \else $\text{FWHM}_{\rm int}^{1.2}$\fi}

\newcommand{\Mbh   }{\ifmmode M_{\text{BH} } \else $M_{\text{BH} }$\fi}
\newcommand{\mbh   }{\ifmmode M_{\text{BH} } \else $M_{\text{BH} }$\fi}
\newcommand{\MSEE   }{\ifmmode M_{\rm BH}^{\rm SE}  \else $M_{\rm BH}^{\rm SE}$\fi}
\newcommand{\MSE   }{\ifmmode M_{\rm BH}^{\rm SE}\left(\fwhm, L_{\lambda}\right)  \else $M_{\rm BH}^{\rm SE}\left(\fwhm, L_{\lambda}\right)$\fi}
\newcommand{\MSIGMA  }{\ifmmode M_{\rm BH}^{\rm \sigma^{\star}}  \else $M_{\rm BH}^{\rm \sigma^{\star}}$\fi}
\newcommand{\MAD   }{\ifmmode M_{\rm BH}^{\rm AD}  \else $M_{\rm BH}^{\rm AD}$\fi}
\newcommand{\MSEcor }{\ifmmode M_{\rm BH}^{\rm SE}(\rm corr)  \else $M_{\rm BH}^{\rm SE}(\rm corr)$\fi}
\newcommand{\MADfm   }{\ifmmode M_{\rm BH}^{\rm AD}\left(\MAD,\fwhm\right)  \else $M_{\rm BH}^{\rm AD}\left(\MAD,\fwhm\right)$\fi}
\newcommand{\MADfl   }{\ifmmode M_{\rm BH}^{\rm AD}\left(\fwhm, L_{\lambda}\right)  \else $M_{\rm BH}^{\rm AD}\left(\fwhm, L_{\lambda}\right)$\fi}
\newcommand{\deltaM   }{\ifmmode \Delta \log M_{\text{BH} } \else $\Delta \log M_{\text{BH} }$\fi}
\newcommand{\fad   }{\ifmmode   f_{\rm AD  } \else $  f_{\rm AD  }$\fi}
\newcommand{\fadl   }{\ifmmode  f_{\rm AD  }\left({\rm line}\right) \else $  f_{\rm AD  }\left({\rm line}\right)$\fi}

\newcommand{\Mdot  }{\ifmmode \dot{M} \else $\dot{M}$\fi}
\newcommand{\mdot  }{\ifmmode \dot{m} \else $\dot{M}$\fi}
\newcommand{\LLedd }{\ifmmode L/L_{\text{Edd} } \else $L/L_{\text{Edd} }$\fi}
\newcommand{\lledd }{\ifmmode L/L_{\text{Edd} } \else $L/L_{\text{Edd} }$\fi}

\newcommand{\astar }{\ifmmode a_{*} \else  $a_{*}$\fi}
\newcommand{\RBLR  }{\ifmmode R_{\text{BLR} } \else $R_{\text{BLR}}$\fi}

\newcommand{  \Halpha   }{\ifmmode {\text{H} }\alpha \else H$\alpha$\fi}
\newcommand{  \ha   	}{\ifmmode {\text{H}}\alpha \else H$\alpha$\fi}
\newcommand{  \Hbeta    }{\ifmmode {\text{H} }\beta \else H$\beta$\fi}
\newcommand{  \hb    	}{\ifmmode {\text{H} }\beta \else H$\beta$\fi}
\newcommand{  \mgii     }{\ifmmode {\text{Mg} }\,\textsc{ii} \else Mg\,\textsc{ii}\fi}
\newcommand{  \MgII    }{\ifmmode {\text{Mg} }\,\textsc{ii}\,\lambda2798 \else Mg\,\textsc{ii}\,$\lambda2798$\fi}
\newcommand{  \HeII    }{\ifmmode {\text{He} }\,\textsc{ii}\,\lambda1640 \else He\,\textsc{ii}\,$\lambda1640$\fi}
\newcommand{  \NIII     }{\ifmmode {\text{N} }\,\textsc{iii}]\,\lambda1750 \else N\,\textsc{iii}]\,$\lambda1750$\fi}
\newcommand{  \NIV     }{\ifmmode {\text{N} }\,\textsc{iv}\,\lambda1718 \else N\,\textsc{iv}\,$\lambda1718$\fi}
\newcommand{  \NIVo     }{\ifmmode {\text{N} }\,\textsc{iv}]\,\lambda1486 \else N\,\textsc{iv}]\,$\lambda1486$\fi}
\newcommand{  \OIII     }{\ifmmode {\text{O} }\,\textsc{iii}]\,\lambda1663 \else O\,\textsc{iii}]\,$\lambda1663$\fi}
\newcommand{  \oiii     }{\ifmmode {\text{O} }\,\textsc{iii}]\,\lambda5007 \else O\,\textsc{iii}]\,$\lambda5007$\fi}
\newcommand{  \civ      }{\ifmmode {\text{C} }\,\textsc{iv}  \else C\,\textsc{iv}\fi}
\newcommand{  \ciii      }{\ifmmode {\text{C} }\,\textsc{iii}]  \else C\,\textsc{iii}]\fi}
\newcommand{  \CIV      }{\ifmmode {\text{C} }\,\textsc{iv}\,\lambda1549 \else C\,\textsc{iv}\,$\lambda1549$\fi}
\newcommand{  \CIII      }{\ifmmode {\text{C} }\,\textsc{iii}]\,\lambda1909 \else C\,\textsc{iii}]\,$\lambda1909$\fi}
\newcommand{  \SiOIV      }{\ifmmode {\text{Si} }\,\textsc{IV}+{\text{O} }\,\textsc{IV}]\, \lambda1400 \else Si\,\textsc{iv}+O\,\textsc{iv}]  \,$\lambda1400$\fi}
\newcommand{  \sioiv      }{\ifmmode {\text{Si} }\,\textsc{IV}+{\text{O} }\,\textsc{IV}]\ \else Si\,\textsc{iv}+O\,\textsc{iv}]\fi}
\newcommand{  \feii      }{\ifmmode {\text{Fe} }\,\textsc{ii}  \else Fe\,\textsc{ii}\fi}

\newcommand{  \FeII      }{\ifmmode {\text{Fe} }\,\textsc{ii}  \else Fe\,\textsc{ii}\fi}

\newcommand{  \FeIII      }{\ifmmode {\text{Fe} }\,\textsc{iii}  \else Fe\,\textsc{iii}\fi}

\newcommand{  \Luv      }{\ifmmode L_{1450} \else $L_{1450}$\fi}
\newcommand{  \Lop      }{\ifmmode L_{5100} \else $L_{5100}$\fi}
\newcommand{  \Loploc   }{\ifmmode L_{5100}^{\text{local}} \else $L_{5100}^{\text{local}}$\fi}
\newcommand{  \Lopglob  }{\ifmmode L_{5100}^{\text{global}} \else $L_{5100}^{\text{global}}$\fi}
\newcommand{  \Lsix     }{\ifmmode L_{6200} \else $L_{6200}$\fi}
\newcommand{  \Lthree   }{\ifmmode L_{3000} \else $L_{3000}$\fi}

\newcommand{ \Lhb   }{\ifmmode L\left(\hb\right) \else $L\left(\hb\right)$\fi}
\newcommand{ \fwhb  }{\ifmmode {\text{FWHM}_{\rm obs} }\left(\hb\right) \else $\text{FWHM}_{\text{obs}}(\hb)$\fi}

\newcommand{ \fwhbloc  }{\ifmmode {\text{FWHM} }\left(\hb\right)_{\text{local} } \else FWHM(\hb)_{\text{local} } \fi}
\newcommand{ \fwhbglob  }{\ifmmode {\text{FWHM} }\left(\hb\right)_{\text{global} } \else FWHM(\hb)_{\text{global} } \fi}
\newcommand{\deltaMhb   }{\ifmmode \Delta \log M_{\rm BH}\left({\rm \Hbeta}\right)  \else $\Delta \log M_{\rm BH}\left({\rm \Hbeta}\right)$\fi}
\newcommand{\MSEhb   }{\ifmmode M_{\rm BH}^{\rm SE}\left({\rm \Hbeta}, \Lop \right)  \else $M_{\rm BH}^{\rm SE}\left({\rm \Hbeta}, \Lop \right)$\fi}

\newcommand{ \fwha  }{\ifmmode {\text{FWHM}_{\rm obs} }\left(\ha\right) \else $\text{FWHM}_{\text{obs}}(\ha)$\fi}
\newcommand{\deltaMha   }{\ifmmode \Delta \log M_{\rm BH}\left({\rm \Halpha}\right) \else $\Delta \log M_{\rm BH}\left({\rm \Halpha}\right)$\fi}
\newcommand{\MSEha   }{\ifmmode M_{\rm BH}^{\rm SE}\left({\rm \Halpha}, \Lsix \right)  \else $M_{\rm BH}^{\rm SE}\left({\rm \Halpha}, \Lsix \right)$\fi}

\newcommand{ \Lmg   }{\ifmmode L\left(\mgii\right) \else $L\left(\mgii\right)$\fi}
\newcommand{ \fwmg  }{\ifmmode {\text{FWHM}_{\rm obs} }\left(\mgii\right) \else $\text{FWHM}_{\text{obs}}(\mgii)$\fi}

\newcommand{\deltaMmg   }{\ifmmode \Delta \log M_{\rm BH}\left({\rm \mgii}\right)  \else $\Delta \log M_{\rm BH}\left({\rm \mgii}\right)$\fi}
\newcommand{\MSEmg   }{\ifmmode M_{\rm BH}^{\rm SE}\left({\rm \mgii}, \Lthree \right)  \else $M_{\rm BH}^{\rm SE}\left({\rm \mgii}, \Lthree \right)$\fi}

\newcommand{ \Lciv  }{\ifmmode L\left(\civ\right) \else $L\left(\civ\right)$\fi}

\newcommand{ \fwciv  }{\ifmmode {\text{FWHM}_{\rm obs} }\left(\civ\right) \else $\text{FWHM}_{\text{obs}}(\civ)$\fi}
\newcommand{\deltaMciv   }{\ifmmode \Delta \log M_{\rm BH}\left({\rm \civ}\right)  \else $\Delta \log M_{\rm BH}\left({\rm \civ}\right)$\fi}
\newcommand{\MSEciv   }{\ifmmode M_{\rm BH}^{\rm SE}\left({\rm \civ}, \Luv \right)  \else $M_{\rm BH}^{\rm SE}\left({\rm \civ}, \Luv \right)$\fi}

\newcommand{ \voff  }{\ifmmode v_{\text{off} } \else $v_{\text{off} }$\fi} 

\newcommand{ \sigline  }{\ifmmode \sigma_{\text{line}} \else $\sigma_{\text{line}}$\fi}

\newcommand{ \sigmamg  }{\ifmmode {\sigma }\left(\mgii\right) \else $\sigma$(\mgii)\fi}
\newcommand{ \sigmaciv  }{\ifmmode {\sigma }\left(\civ\right) \else $\sigma$(\civ)\fi}
\newcommand{ \sigmahb  }{\ifmmode {\sigma }\left(\Hbeta\right) \else $\sigma$(\Hbeta)\fi}
\newcommand{ \sigmaha  }{\ifmmode {\sigma }\left(\Halpha\right) \else $\sigma$(\Halpha)\fi}

\newcommand{\MbhHb   }{\ifmmode M_{\text{BH}  } \left( \Hbeta \right) \else $M_{\text{BH} } \left( \Hbeta \right)$\fi}
\newcommand{\MbhHa   }{\ifmmode M_{\text{BH} }  \left( \Halpha \right) \else $M_{\text{BH} } \left( \Halpha \right)$\fi}%
\newcommand{\MbhMg   }{\ifmmode M_{\text{BH}  } \left( \mgii \right) \else $M_{\text{BH} } \left( \mgii \right)$\fi}
\newcommand{\MbhC   }{\ifmmode M_{\text{BH}  } \left( \civ \right) \else $M_{\text{BH} } \left( \civ \right)$\fi}

\newcommand{\Mbhfw   }{\ifmmode M_{\text{BH}}\left(\text{FWHM}\right)    \else $M_{\text{BH}}\left(\text{FWHM}\right)$\fi}
\newcommand{\Mbhsig   }{\ifmmode M_{\text{BH}}\left(\sigma_{\text{line}}\right)    \else $M_{\text{BH}}\left(\sigma_{\text{line}}\right)$\fi}
\newcommand{\fadha  }{\ifmmode   f_{\rm AD  }\left({\rm \Halpha}\right) \else $  f_{\rm AD  }\left({\rm \Halpha}\right)$\fi}
\newcommand{\fadhb  }{\ifmmode   f_{\rm AD  }\left({\rm \Hbeta}\right) \else $  f_{\rm AD  }\left({\rm \Hbeta}\right)$\fi}
\newcommand{\fadmgii  }{\ifmmode   f_{\rm AD  }\left({\rm MgII}\right) \else $  f_{\rm AD  }\left({\rm MgII}\right)$\fi}

\newcommand{\fw}{\ifmmode \fwhm_{\text{local}} \else $\fwhm_{\text{local}}$\fi}
\newcommand{\fwtdc   }{\ifmmode \fwhm_{\text{global}} \else $\fwhm_{\text{global}}$\fi}
\newcommand{\Llocal   }{\ifmmode L_{\text{local}} \else $L_{\text{local}}$\fi}
\newcommand{\Ltdc   }{\ifmmode L_{\text{global}} \else $L_{\text{global}}$\fi}

\newcommand{ \mumg  }{\ifmmode \mu\left(\mgii\right) \else $\mu\left(\mgii\right)$\fi}
\newcommand{ \fmg   }{\ifmmode f\left(\mgii\right) \else $f\left(\mgii\right)$\fi}
\newcommand{ \muciv }{\ifmmode \mu\left(\civ\right) \else $\mu\left(\civ\right)$\fi}
\newcommand{ \fciv  }{\ifmmode f\left(\civ\right) \else $f\left(\civ\right)$\fi}

\newcommand{\Ntot}{39}

\newcommand{\ergs}	{\ifmmode {\text{erg\,s}}^{-1} \else erg s$^{-1}$\fi}
\newcommand{\kms}	{\ifmmode {\text{km\,s} }^{-1} \else km\,s$^{-1}$\fi}
\newcommand{\todo}{\ifmmode \text{\Huge{\(\bullet\)}} \else {\Huge$\bullet$}\fi}

\title{The effect of nuclear gas distribution on the mass determination of supermassive black holes}

\author{J. E. Mejia-Restrepo$^{1}$, P. Lira$^{1}$, H. Netzer$^{2}$, B. Trakhtenbrot$^{3}$, D. M. Capellupo$^{4}$ }

\begin{document}

\maketitle

\begin{affiliations}
\item Departamento de Astronom\'{i}a, Universidad de Chile, Camino el Observatorio 1515, Santiago, Chile.
\item School of Physics and Astronomy, Tel Aviv University, Tel Aviv 69978, Israel. 
\item Institute for Astronomy, Dept. of Physics, ETH Zurich, Wolfgang-Pauli-Strasse 27, CH-8093 Zurich, Switzerland.
\item Department of Physics, McGill University, Montreal, Quebec, H3A 2T8, Canada.
\end{affiliations}

\begin{abstract}
Supermassive black holes  reside in the nuclei of most galaxies. Accurately determining their mass  is key to understand how the population evolves over time and how the black holes relate to their host galaxies\cite{Ferrarese2000,Xiao2011,KormendyHo2013}. Beyond the local universe,  the  mass is commonly estimated assuming virialized motion of gas in the close vicinity to the active black holes, traced through broad emission lines\cite{TrakhtenbrotNetzer2012,Shen2013}. However, this procedure has uncertainties associated with the unknown distribution of the gas clouds. 
Here we show that the comparison of black hole masses derived from the properties of the central accretion disc with the virial mass estimate provides 
a correcting factor, for the virial mass estimations, that is inversely proportional to the observed width of the broad emission lines. Our results suggest that line-of-sight inclination of gas  in a planar distribution can account for this effect. However, radiation pressure effects on the distribution of gas  can also reproduce our findings. Regardless of the physical origin, our findings contribute to mitigate the uncertainties in current  black hole mass estimations and, in turn, will  help to further understand the evolution of distant supermassive black holes and their host galaxies.
\end{abstract}

Active Supermassive black holes (SMBHs) are powered by accretion flows, probably in the form of accretion discs 
(ADs) that convert gravitational energy into radiation\cite{ShakuraSunyaev1973}. 
Gas in the Broad Line Region (BLR), located in the vicinity  of the SMBH and moving 
at Keplerian velocities of thousands of kilometres per second, is photo-ionized by the 
AD producing broad emission lines. Under virial equilibrium, the observed width of these lines (in terms of full width at half maximum, \fwobs) can be 
used as a proxy for the virial velocity ($V_{\text{BLR}}$) and \Mbh\ can be expressed as:
\begin{equation}
 \Mbh =  G^{ - 1} R_{\text{BLR} }\ V_{\text{BLR}}^{2}  = f\ G^{ - 1} R_{\text{BLR} }\ \fwhm_{\rm{obs}}^{2} 
 \label{eqn:virial_eqn}
\end{equation}
Here, $G$ is the gravitational constant, $R_{\text{BLR}}$ is the mean BLR distance to 
the SMBH and $f$ is the virial factor that accounts for the differences between the unknown 
$V_{\text{BLR}}$ and \fwobs\ that are mostly caused by the BLR gas distribution of each object. Since even in the closest active galaxies the BLR cannot be 
resolved with current capabilities, $R_{\text{BLR}}$ is estimated from reverberation mapping 
(RM) experiments that show a strong correlation between the typical distance to the \Hbeta\ emitting 
region and the continuum luminosity (the $R_{\text{BLR}}-L\,$ relation)\cite{Kaspi2000,Bentz2013}. 
$f$ is assumed to be constant for all systems and is usually determined by requiring 
RM-based masses (from Equation \ref{eqn:virial_eqn}) to agree, on average, with masses 
estimated from the relation between \Mbh\ and the stellar velocity dispersion found in local 
galaxies\cite{Onken2004, Graham2015, Woo2015}. This indirect technique to determine 
\Mbh\ is known as the ``single epoch virial method''\cite{TrakhtenbrotNetzer2012,Shen2013}.

\begin{figure}[!ht]
 \includegraphics[width=0.96\columnwidth,keepaspectratio]{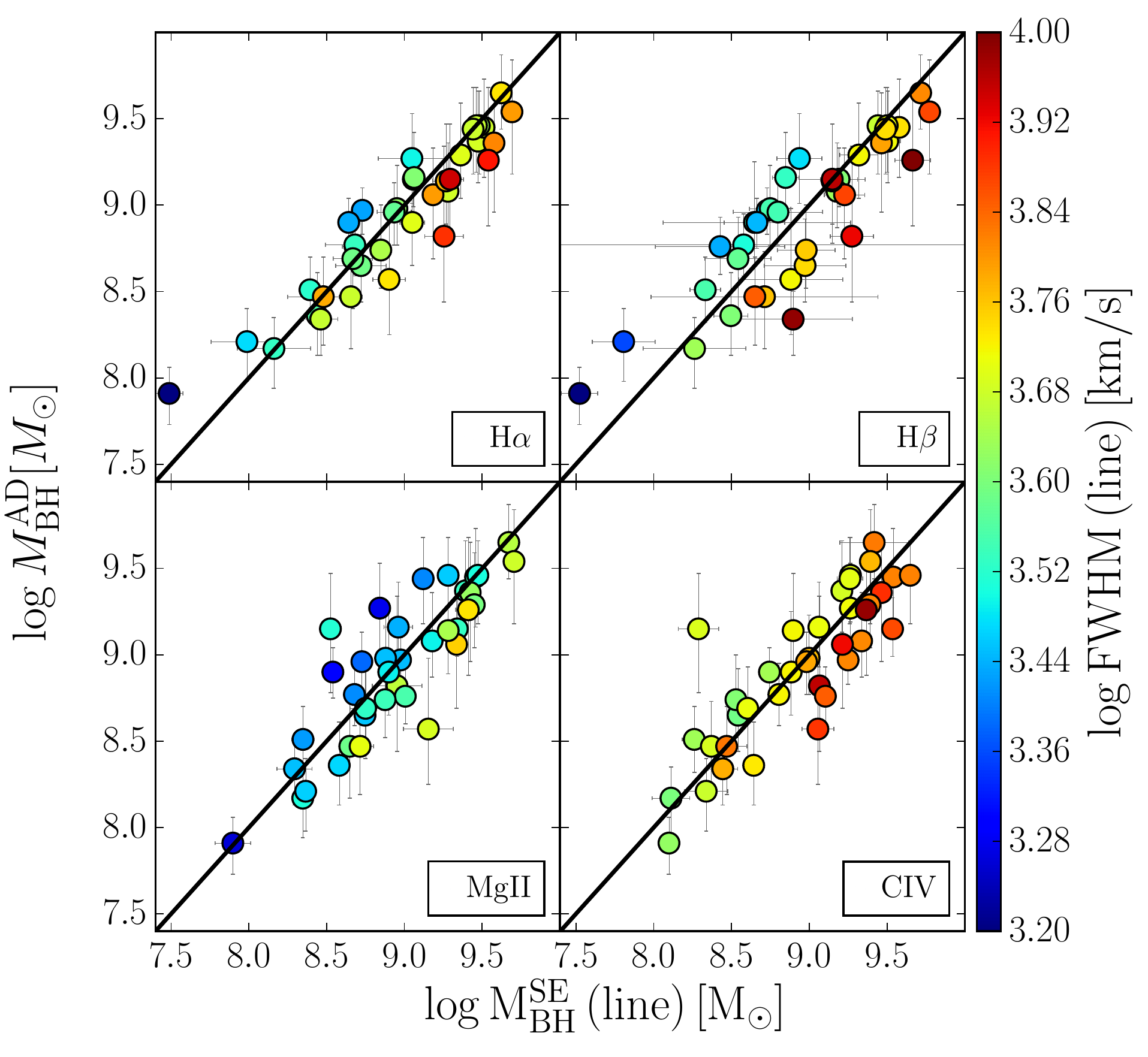}
 \caption{ \textbf{ Comparison of the accretion-disk based and single-epoch based black hole mass determinations}.  Accretion-disc-derived black hole masses versus single  epoch black hole masses. The black solid line shows the 1:1
   relation.  The colour of the points scales with the \fwobs\ of the emission lines in 
   each panel. 
   The approaches yield masses in very good agreement with each other,
   albeit with a significant scatter. The scatter shows a strong gradient with \fwobs, 
   from small \fwobs\ (blue) above the 1:1 relation to large \fwobs\ (red) below. The \MAD\  error bars   enclose the central 68\% of its marginalized posterior probability distribution  and the \MSEE\  error bars enclose  the central 68\% of the probability distribution after 100 Monte-Carlo realizations of the spectral fitting procedure. }
 \label{fig:f1}
\end{figure}

Unfortunately, the virial method is subject to biases and uncertainties associated with
our ignorance of the dependence of $f$ on additional physical properties.  
These could include radiation pressure 
perturbations\cite{Marconi2008,NetzerMarziani2010}, non virial velocity
components\cite{Denney2009, Denney2010}, the relative thickness ($H/R_{\rm BLR}$) of the
Keplerian BLR orbital plane\cite{Gaskell2009}, and the line-of-sight inclination angle
($i$)\cite{Wills1986,ShenHo2014,Runnoe2014} of this plane.
An analytical expression for $f$ in the case of a planar BLR of thickness
$H/R_{\rm BLR}$ is given by:
\begin{equation}
 f = \left[4   \left(\sin^2 i + \left(H/R_{\rm BLR}\right)^2 \right)\right]^{-1}
 \label{eqn:virial_factor}
\end{equation}
where $\sin^2 i$ accounts for the line-of-sight projection of the  
Keplerian velocity of the BLR orbital  plane\cite{Collin2006,Decarli2008a}. The nature of the velocity
component responsible for the thickness of 
the BLR in unclear. However, ideas such as non-coplanar orbits, accretion disc radiation pressure,
induced turbulence and outflowing disc winds have been suggested in the
literature as plausible mechanisms to puff up the BLR\cite{Collin2006,Czerny2016}. 
Given all these, the assumption of an universal $f$  introduces 
an uncertainty in the single epoch method which is estimated to be at least a factor 
of 2-3.

Recently, we used an alternative method to estimate \Mbh\ by fitting the AD spectra of 
37 active galaxies at z$\sim$1.5 (about 1/3 of the 
current age of the universe), observed using the ESO X-Shooter spectrograph which
provides simultaneous, very wide wavelength coverage of the AD
emission\cite{Capellupo2015,Capellupo2016} (see supplementary
information for sample description).  The spectra were fitted with standard,
geometrically thin, optically thick AD models\cite{ShakuraSunyaev1973} including general
relativistic and disc atmosphere corrections\cite{SloneNetzer2012}. In our modelling we made sure to avoid known model uncertainties that affect the short wavelength region ($\lambda<1216$\AA, see supplementary information). Each model is 
determined by several properties, mainly
 its \Mbh\ (\MAD), the normalized accretion rate (expressed as the Eddington ratio $\lambda_{\rm Edd} = L/L_{\rm Edd}$), the black hole spin (\astar)  and the disc inclination to the line of sight (see supplementary
information for model description).  Crucially, this method only relies
on our ability to model the AD. As a result, the derived masses are 
independent of the  BLR geometry and kinematics,
and therefore of any assumptions on the $f$ factor.  For this sample,
we also previously estimated the associated single epoch black hole masses (\MSEE) from
the \Halpha, \Hbeta, \mgii\ and \civ\ broad emission lines\cite{MejiaRestrepo2016a}. As the $R_{\text{BLR}}-L\,$ 
relation has only been robustly  established for the \Hbeta\ 
line, the \Halpha, \mgii and \civ\ single epoch masses are cross-calibrated to agree on average 
with the \Hbeta\ mass estimations.

\Mbh\ determinations from these two methods are compared in Figure \ref{fig:f1}.
The approaches yield masses in very good agreement with each other (suggesting that AD and SE masses have comparable accuracies)
albeit with significant scatter of a factor of about two\cite{Capellupo2016}.  In this letter we
 looked for  possible drivers  for this scatter and found a strong gradient in \fwobs\ 
across the relation, as can be seen by the colour gradient of the data points in Figure
\ref{fig:f1}.

\begin{figure}
 \includegraphics[width=0.5\textwidth]{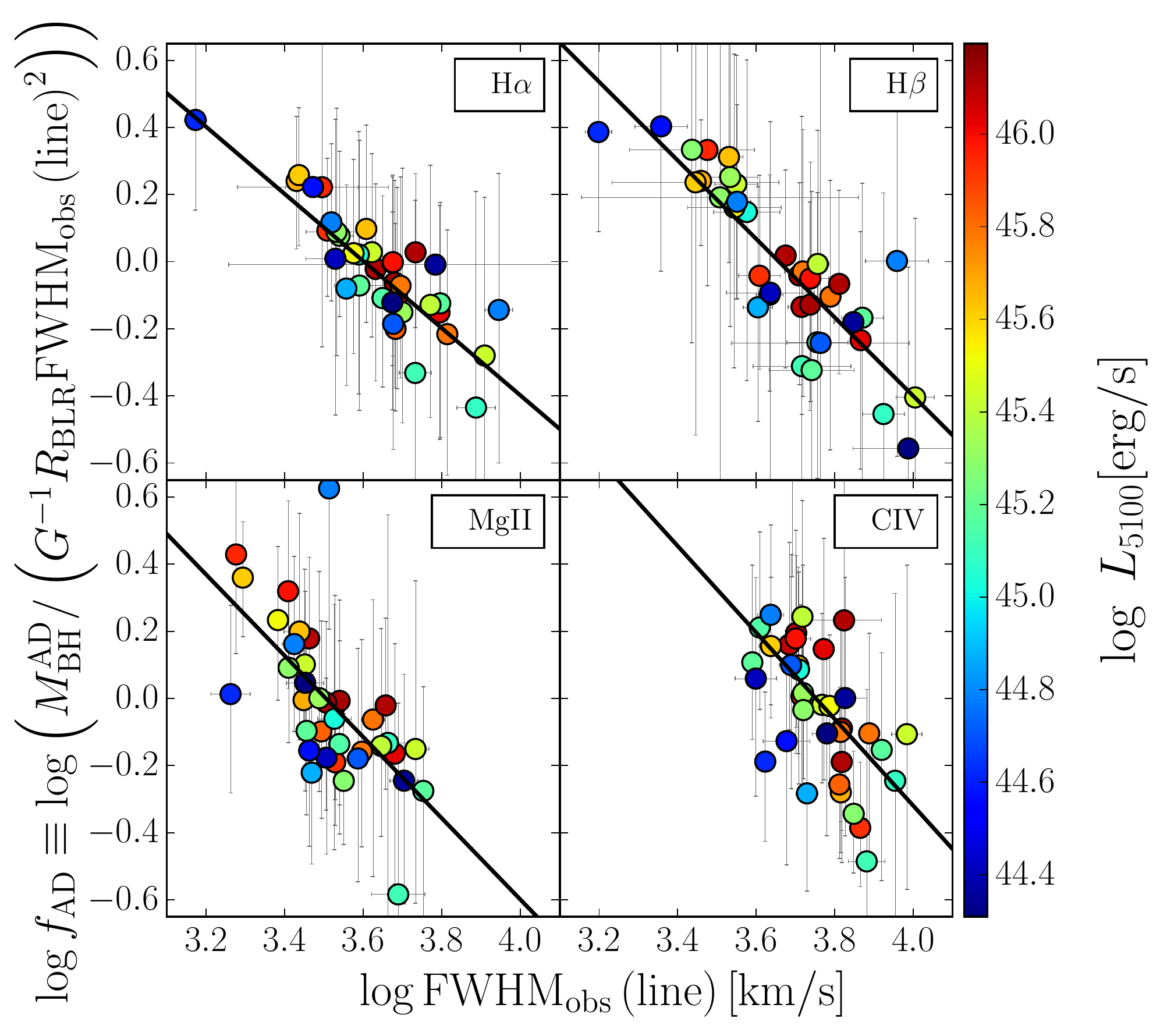}
 \caption{ \textbf{Virial factor $f$ as a function of \fwobs\ for the \Halpha, \Hbeta,
  \mgii\ and \civ\ broad emission lines.} The black solid line is the best linear fit to 
  the data. There is a clear anti-correlation between \fad\ and \fwobs\ for all 
  lines as suggested in Figure \ref{fig:f1}. The colour of the points scales with the measured 
  monochromatic luminosity at 5100\AA\ ($L_{5100}$) for each object, as indicated by the
 colour bar. Redder (bluer) points correspond to larger (smaller) values of $L_{5100}$.
 No clear gradient is seen in the scatter of these anti-correlations as a function of \Lop.  The \fwobs\ error bars  enclose  the central 68\% of the probability distribution after 100 Monte-Carlo realizations of the spectral fitting procedure and the \fad\ error bars  are estimated  from error propagation of the quantities involved on the calculation of this quantity.}
 \label{fig:f2}
\end{figure}

The ratio between \MAD\ and $\MSEE/f=G^{ - 1}R_{\text{BLR} }\fwobstwo$
allows us to determine a proxy for the virial factor $f$ which we define
as $f_{\rm AD}\left({\rm line}\right) \equiv M_{\rm BH}^{\rm AD}\ /
\left( G^{ - 1} R_{\rm BLR}\ \left(\fwobs \left({\rm line}\right)\right)^2
\right)$.  In Figure \ref{fig:f2} 
we show \fadl\ as a function of the \fwobs\ for the \Halpha, \Hbeta, \mgii\ 
and \civ\ broad emission lines. Strong anti-correlations between \fad\ and \fwobs\ 
are present for all lines. These correlations are found to be significantly
stronger than the expected correlations between \fad\ and 
$G^{-1}R_{\text{BLR} }\fwobstwo$ (see Table \ref{tab:t1} and see supplementary 
information  for details).  We can conclude that 
the \fwobs\ of the broad lines drives the discrepancies between \MAD\ and \MSEE. 

\begin{table*}[!ht]
\centering
\fontsize{9}{5}
\resizebox{\textwidth}{!}{
\begin{tabular}{ccccccc}
\hline 
& & & \multicolumn{2}{c}{\fwobs (\dag)}  & \multicolumn{2}{c}{$G^{ - 1}R_{\text{BLR} }\fwobstwo$ (\ddag)}\\
      Broad line  & ${\rm FWHM}^0_{\rm obs}\left[\kms\right]$ & $\beta$     & r$_{\rm s}$ & P$_{\rm s}$   & r$_{\rm s}$ & P$_{\rm s}$\\ 
      \hline
\Halpha & 4000$\pm$700                 & -1.00$\pm$0.10 & -0.85 & 4$\times 10^{-11}$ & -0.44 & 5$\times 10^{-3}$  \\
\Hbeta  & 4550$\pm$1000                & -1.17$\pm$0.11 & -0.84 & 8$\times 10^{-11}$ & -0.48 & 2$\times 10^{-3}$ \\
\MgII   & 3200$\pm$800                 & -1.21$\pm$0.24 & -0.75 & 9$\times 10^{-8}$ & -0.23 & 2$\times 10^{-1}$ \\
\CIV    & 5650$\pm$3000                & -1.29$\pm$0.35 & -0.61&  6$\times 10^{-5}$ & -0.25 & 1$\times 10^{-1}$ \\ \hline  
\end{tabular}
}
\caption{\textbf{The virial factor as a function of \fwobs\ for the broad emission lines.} 
${\rm FWHM}^0_{\rm obs}$ and $\beta$ are best fit parameters found for 
$\fad=\left(\fwobs\left(\text{line}\right)/ {\rm FWHM}^0_{\rm obs}\right)^{\beta}$.
r$_{\rm s}$ and P$_{\rm s}$ are the Spearman correlation coefficient and 
associated null-hypothesis probability for the \fad\ vs \fwobs\ (\dag) and
\fad\ vs $G^{ - 1}R_{\text{BLR} }\fwobstwo$ (\ddag) correlations.}
\label{tab:t1}
\end{table*}

We also determined how \MAD\ depends on \fwobs\ and $L_{\lambda}$ (see supplementary
information).  The dependence on \fwobs\ is close to linear and, therefore, very 
different from the expected squared dependency found in Equation \ref{eqn:virial_eqn}.  
The dependence on the monochromatic luminosities is consistent, within errors, with 
that found for single epoch calibrations (i.e., $\MAD \propto R_{\rm BLR}$). 
This indicates that $L_{\rm \lambda}$ has no impact on the scatter between \MSEE\ and
\MAD\ and that $f$ can be expressed as a single function of the \fwobs\ of the broad 
emission lines. As can be seen  in Table \ref{tab:t1}, our measurements are consistent
within uncertainties with $\fad \propto \fwobs^{-1}$ for all lines.  

Previous works attempted to derive $f$ by comparing single epoch SMBH
mass estimations with masses obtained from  alternative methods. 
For instance,  from the scaling relations
between the black hole mass and the  luminosity\cite{Decarli2008b} or the
stellar dispersion\cite{ShenHo2014} of the host galaxy spheroidal
components as well as from the amplitude of the excess X-ray variability  variance that is found to be inversely anti-correlated with the black hole mass\cite{Nikolajuk2006}. The results of these works also exhibit an
anti-correlation between $f$ and the \fwobs\ of the broad emission
lines and were understood as an effect of line of sight inclination of the BLR.
However,  these works applies the same prescription to all systems,
assuming that all objects are well represented by the median trend of the scaling 
relations, and do not take into account the large intrinsic scatter in such relations. 
This is in contrast with our sample where \MAD\ is independently obtained
for each object through individual spectral fitting of the accretion disc emission.

\begin{figure}[!hb]
  \includegraphics[scale=0.45]{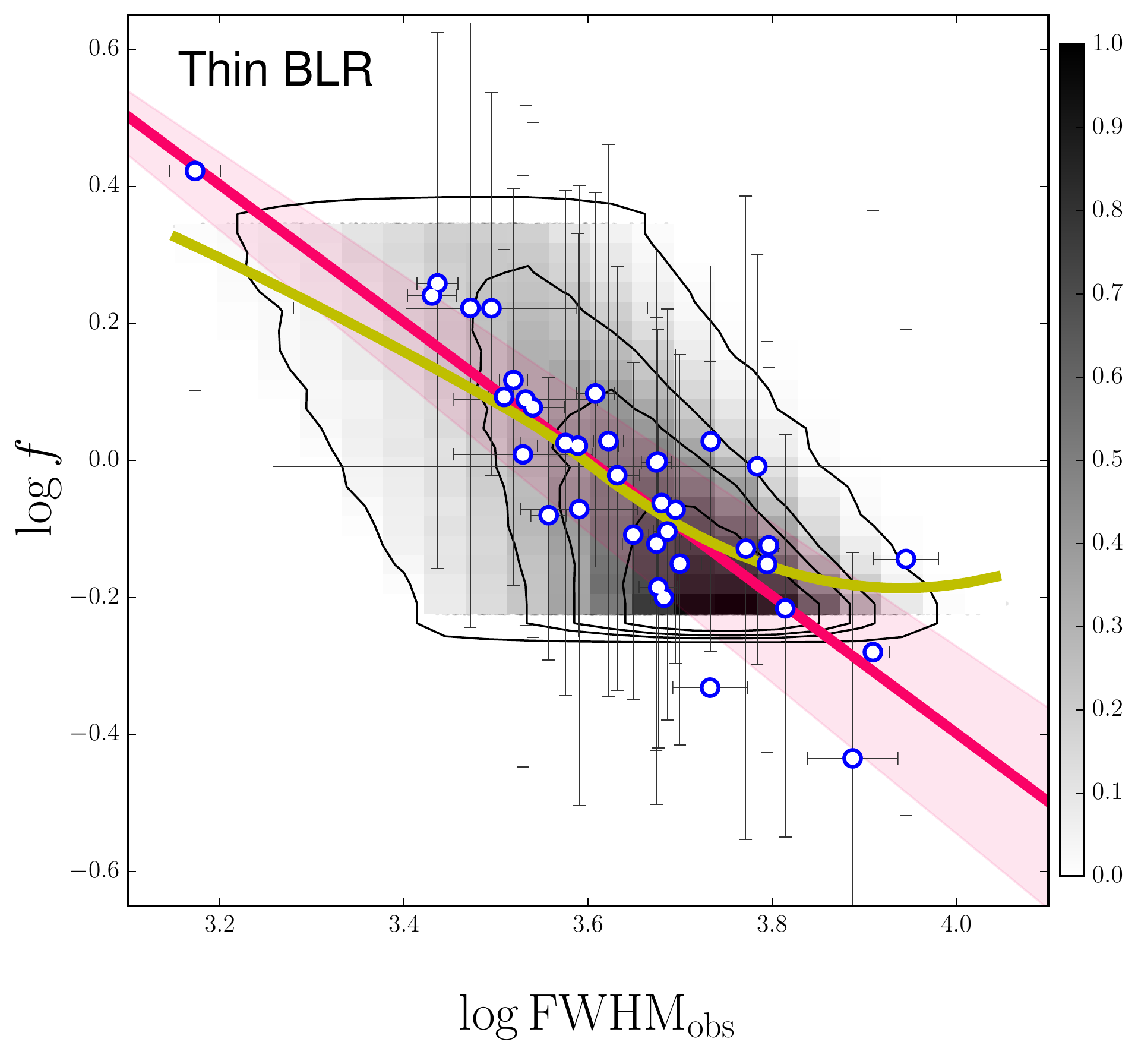}
  \caption{\textbf{Virial factor--\fwobs\ bi-dimensional distribution.}
  Predicted bi-dimensional probability distribution function of the virial factor versus
  \fwobs\ for a thin BLR ($H/R=0$) modified by line-of-sight inclination is shown in gray.  
  The darkest regions represent the most probable combinations of these
  quantities as quantified in the colour bar. The thin, black lines are 
  the 25\%, 50\% and 75\% and 99\% confidence limit contours centred around the maximum 
  probability point. The thick yellow line is the median of the $f$--\fwobs\ 
  distribution derived from a quantile non-parametric spline regression.
  The open-blue circles are data taken from Figure \ref{fig:f2} for the \Halpha\ line.
  The magenta thick line is the derived relation $f = \left(\fwha/4000\ \kms \right)$
  and the shadowed region the associated uncertainties. 
  The yellow and magenta lines are in very good agreement within uncertainties.
  Additionally, the distribution of the data points shows a good agreement 
  with the predicted bi-dimensional distribution confidence limits. 
  Explicitly, we find 21\% of the points inside the central 25\% confidence level region, 
  51\% inside the 50\% confidence level region, 78\% inside the 75\% confidence level region
  and 87\% inside the 99\% confidence level region. The error bars for \fad\ and \fwobs\ are described in the legend of Figure \ref{fig:f2}}
 \label{fig:f3}
 \end{figure}

The high quality spectra in our sample and the careful modelling of its 
broad emission lines allow us to explore in detail whether the 
line of sight inclination in a disc-like BLR can reproduce the observed trends. 
For this purpose we prefer to use the data and correlations determined from the \Halpha\ line
because of its high signal-to-noise ratio\cite{MejiaRestrepo2016a}.
We define \fwint\ as the {\it intrinsic\/} full width at half maximum of the virialized 
velocity component of the BLR. To recover the virial expectation 
$\Mbh \propto \left( \fwint \right)^2$ given by 
Equation \ref{eqn:virial_eqn} we use our result that 
$f\propto \fwobsminus$ for \Halpha\ (or equivalently \MAD\ $\propto$ FWHM$_{\rm obs}$) 
implying that, on average, $\fwint \propto \fwobshalf$. 
First, we adopt a model of a thin BLR (assuming that $H/R\to 0$ in Equation \ref{eqn:virial_factor})
and use Monte Carlo simulations to  find the \fwint\ distribution that, after taking into 
account the line-of-sight inclination effects for randomly orientated BLRs, reproduces the 
cumulative \fwha\ distribution (see supplementary information for further details).
Next, we generate a large population of objects drawn from the \fwint\ distribution
and determine for each of these $f$ and FWHM$_{\rm obs}$. Finally, we compare the 
bi-dimensional $f$--FWHM$_{\rm obs}$ distribution obtained from our data with that 
generated from the simulations. We find that we are able to reproduce not only the mean trend 
of the observed correlation, but also the density distribution of data points, as can be seen 
in Figure \ref{fig:f3}. Furthermore, our simulations can recover the expected $\fwint\propto \fwobshalf$ 
correlation (see extended data Figure \ref{fig:E3}). These results strongly indicate that line-of-sight 
inclination effects cause the observed $f$--FWHM$_{\rm obs}$ anti-correlation.

We also considered the combined effect of inclination and BLR thickness by assuming an universal  
$H/R$  within the range 0-1.  We find that a wide range in thickness ratios ($ H/R\lesssim0.5$) can
reproduce the cumulative distribution function of \fwha, but only relatively thin BLRs 
($ H/R\lesssim0.1$) can reproduce the observed bi-dimensional distribution of \fad\ 
and \fwha, and the predicted $\fwint \propto \fwobshalf$ dependence.

\begin{figure*}[!ht]
\renewcommand\thefigure{E4}
  \includegraphics[width=\textwidth,keepaspectratio]{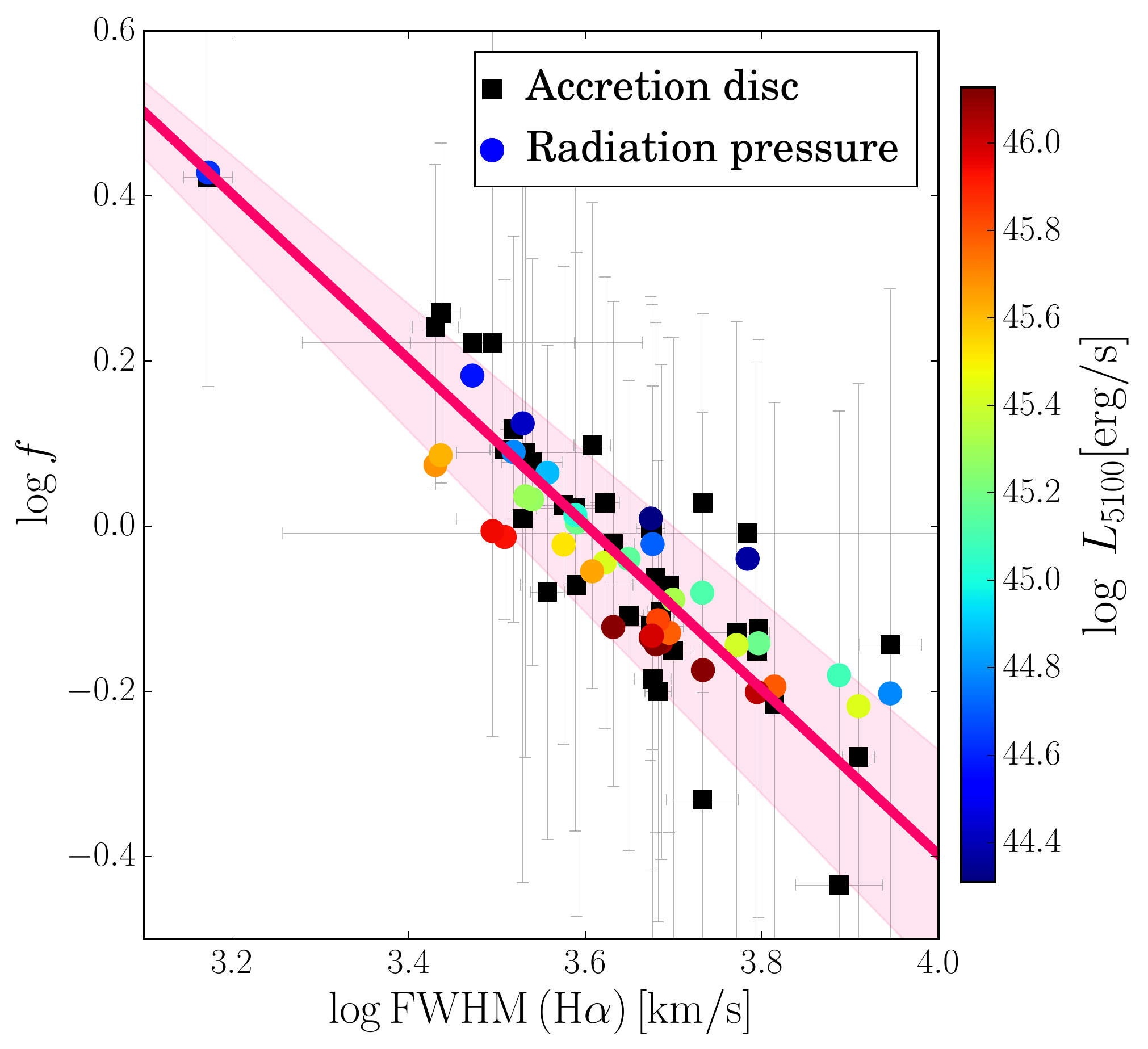}
  \caption{\textbf{Radiation pressure in a gravitationally bound BLR.} The observed virial factor vs \fwobs\
  for the \Halpha\ line is shown (black squares).   The magenta line is the derived relation $f = \left(\fwha/4000\ \kms \right)$ and   the width of the shadowed region accounts for the uncertainties in that relation.  
  The filled points represent the modelled
  $f_{\rm rad}$ from the best fit model for radiation pressure in a gravitationally bound BLR.  The colour of the points scales with the measured 
  monochromatic luminosity at 5100\AA\ ($L_{5100}$) for each object, as indicated by the
 colour bar. Redder (bluer) points correspond to larger (smaller) values of $L_{5100}$.
  As can be observed, the model   predicts that the scatter in $f_{\rm rad}$  (coloured points) is
  driven by  \Lop\ (see Equation \ref{eqn:rad_factor}).   This dependence is  not seen in our data (black squares) as shown in Figure \ref{fig:f2}.  Nevertheless, the relatively large errors in \fad\ and the weak dependence of $f_{\rm rad}$ in \Lop\ may probably hide the expected dependence from this radiation pressure model. The error bars for \fad\ and \fwobs\ are described in the legend of Figure \ref{fig:f2} }
   \label{fig:E4}
 \end{figure*}

We have also examined possible alternative scenarios. In particular,
the effects of radiation pressure force in a gravitationally bound BLR can predict 
$f\propto \fwobsminus$ for some configurations\cite{NetzerMarziani2010}. This model 
predicts that the scatter in the relation will depend on  the luminosity of the 
sources (see supplementary information). However, we do not find 
clear indications for this in our observations, as can be seen by the colour coded data 
points in Figure \ref{fig:f2}, where no clear gradient in \Lop\ is found across the 
$f$--FWHM$_{\rm obs}$ correlation. Note however that given the relatively narrow range 
in \Lop\ covered by our sample (a factor of 80), and the uncertainties in our estimations 
of $f$, radiation pressure remains a intriguing mechanism that should be explored further 
in the future (see Extended data figure \ref{fig:E4}).

Regardless of its physical origin, the dependence  of $f$ on \fwha\ implies that \Mbh\ 
has been, on average, systematically overestimated for systems with large \fwha\ 
($\gtrsim 4000\ \kms$) and underestimated for systems with small \fwha\ ($\lesssim 4000\ \kms$).
The range of \fad\ values presented in Figure \ref{fig:f2}, which are associated with
\fwha$=$1600-8000\ \kms, imply a range in $f$, and hence \Mbh, of factor $\sim$6. However, 
this range should not be taken as representative of the entire population of AGN since 
our sample is too small (37 objects) and was not defined to be complete in terms of 
BLR properties.

Even though our sample is selected at a specific epoch ($z\sim 1.5$),
the physics of a compact region such as the BLR is likely 
to remain constant over time.
This has important implications for the study of active SMBHs at low and high
redshifts.  For example, the lowest \Mbh\ sources at $z\sim 0$ typically show relatively 
narrow BLR profiles (1000\ \kms $\lesssim$\ \fwha$\ \lesssim$\ 2000\ \kms). In this case, 
\Mbh\ should be about 2-4 times larger than current estimates, and consequently 
$\lambda_{\rm Edd}$ should be smaller by the same factor. Another example is related to 
the mass of the most massive young known quasars found at $z \gtrsim 6$. Our proposed 
dependence of $f$ with \fwmg\ reduces the mass of the brightest known systems by up to a 
factor 2, as they typically show lines with \fwmg$\gtrsim$ 3000 \kms,
somewhat alleviating the tension between their outstandingly large masses and the very early
epochs at which they have been discovered \cite{Mortlock2011,Wu2015}




 
 
 



\clearpage
\pagenumbering{arabic}
\setcounter{page}{1}

\begin{figure*}[!ht]
\renewcommand\thefigure{E1}
  \includegraphics[width=0.91\textwidth,keepaspectratio]{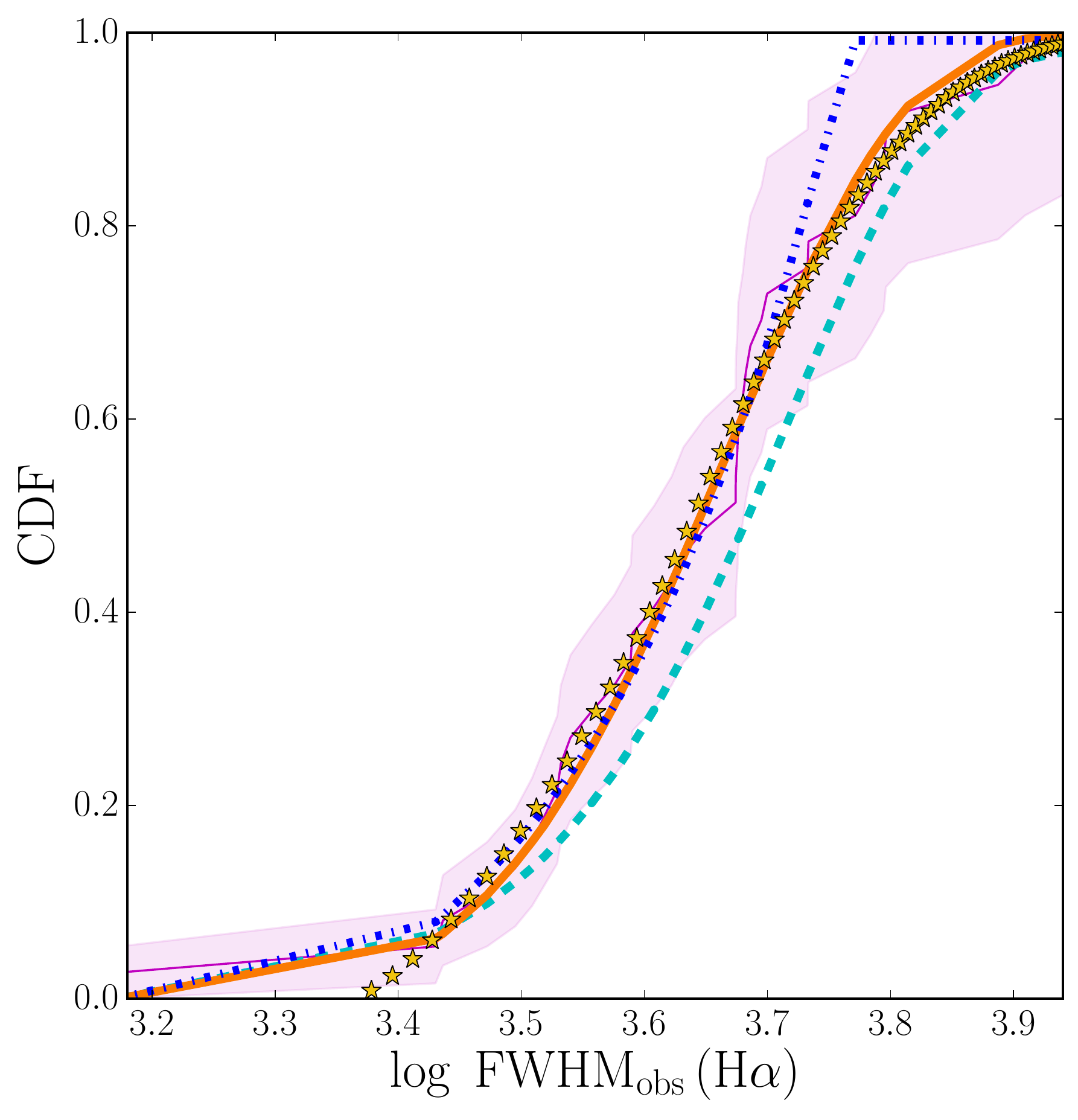}
  \caption{\textbf{Cumulative distribution functions for the observed and simulated \fwobs\ for \Halpha.}
  The cumulative distribution function (CDF) of \fwha\ is the thin magenta line.   
  The magenta shadowed region shows the Poissonian uncertainties. The thick orange 
  line and yellow stars are the modelled CDF for a thin ($H/R=0$) and thick ($H/R=0.5$) 
  BLR, respectively. In both cases we assumed a truncated Gaussian distribution for 
  the intrinsic \fwha\ convolved with a $\sin i$ distribution. The cyan dashed line 
  is a Gaussian distribution with no truncation. The dark blue dashed line is the CDF for
  $\fwint=8170 \kms$ and FWHM$_{\rm std} = 0$, as modelled in other works\cite{MclureDunlop2001, Decarli2008a}. 
  We observe that the modelled CDFs are generally in good agreement with the observed 
  CDF for the thin and thick BLR models, but the thick BLR fails to reproduce the 
  observed CDF at small \fwha\ values.} 
   \label{fig:E1}
 \end{figure*}

\begin{figure*}[!ht]
\renewcommand\thefigure{E2}
  \includegraphics[width=\textwidth,keepaspectratio]{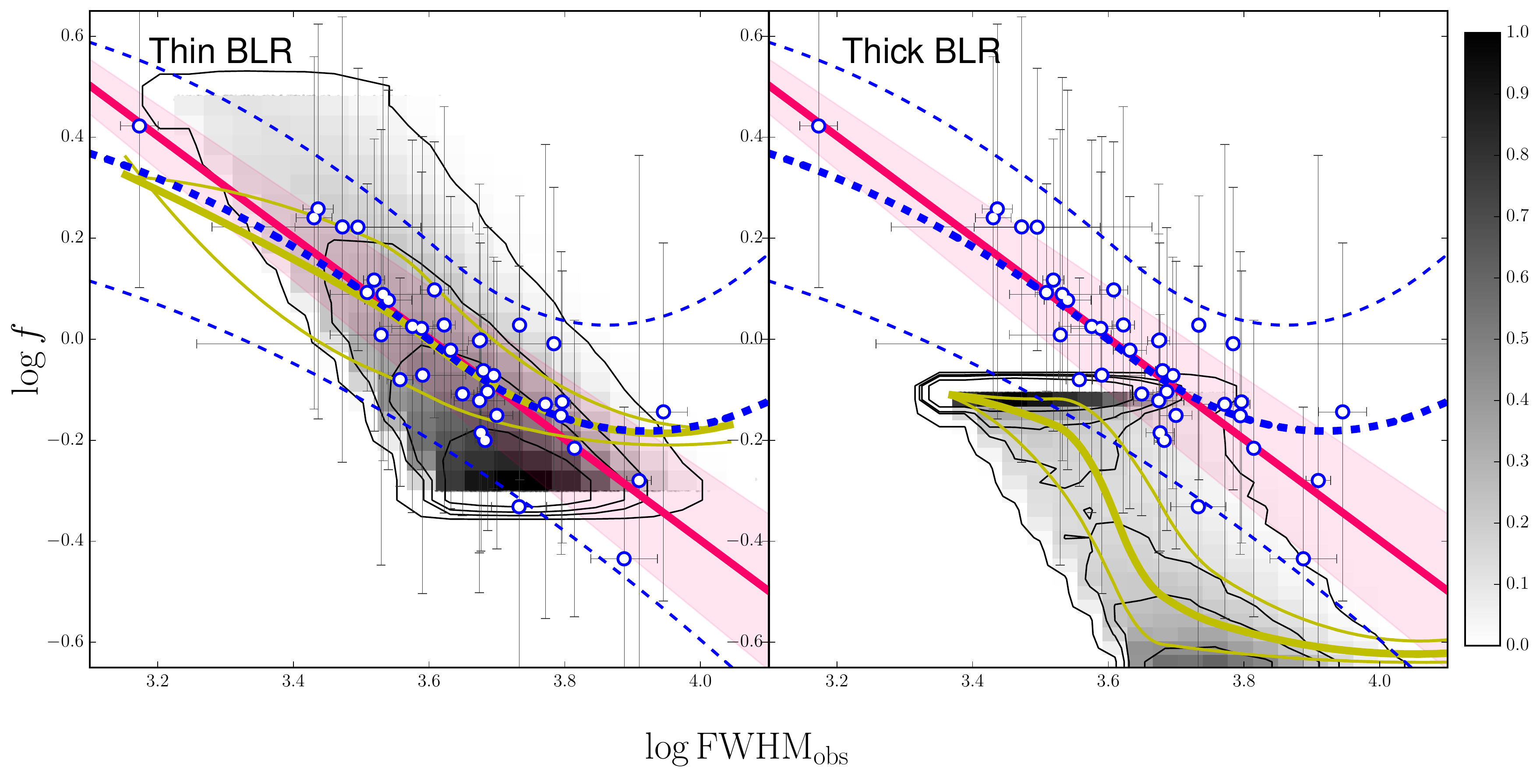}
  \caption{\textbf{Virial factor--\fwobs\ bi--dimensional distributions for a thin and thick BLR.}
  Predicted bi-dimensional probability distribution functions of the virial factor and \fwobs\ for 
  a thin BLR (left) and a thick BLR (right), as predicted by the best-fit models shown in Figure 
  \ref{fig:E1} are shown in gray. The darkest regions represent the most probable combinations of 
  these quantities as quantified in the colour-bar. The thin black lines are the 25\%, 50\% and 75\% 
  and 99\% confidence limit contours centred around the maximum probability point. The thick yellow 
  lines are the median of the $f$--\fwobs\ distributions derived from a quantile non-parametric spline 
  regression. The open-blue circles are data from Figure \ref{fig:f2} for the \Halpha\ line. The magenta 
  lines are the derived relation $f = \left(\fwha/4000\ \kms \right)$ and the shadowed regions the associated 
  uncertainties. The thin blue-dashed lines are the 25\%, 50\% and 75\% quantiles of the observational 
  distribution after accounting for the measurement errors in \fad\ and \fwha. We see that for the thin
  BLR the 50\%-quantile (median) of the theoretical and observational distributions are in very good  
  agreement with each other. Additionally, the distribution of the data points shows good agreement 
  with the predicted bi-dimensional distribution confidence limits. Explicitly, we find that 21\% of 
  the points fall inside the central 25\% confidence level region, 51\% fall inside the 50\% confidence 
  level region, 78\% fall inside the 75\% confidence level region, and 87\% fall inside the 99\% confidence 
  region level. On the other hand, the thick BLR model cannot reproduce the bi-dimensional  $f$--\fwobs\ 
  distribution. The errors bars for \fad\ and \fwobs\ are described in the legend of Figure \ref{fig:f2}} 
   \label{fig:E2}
 \end{figure*}

\begin{figure*}[!ht]
\renewcommand\thefigure{E3}
  \includegraphics[width=\textwidth,keepaspectratio]{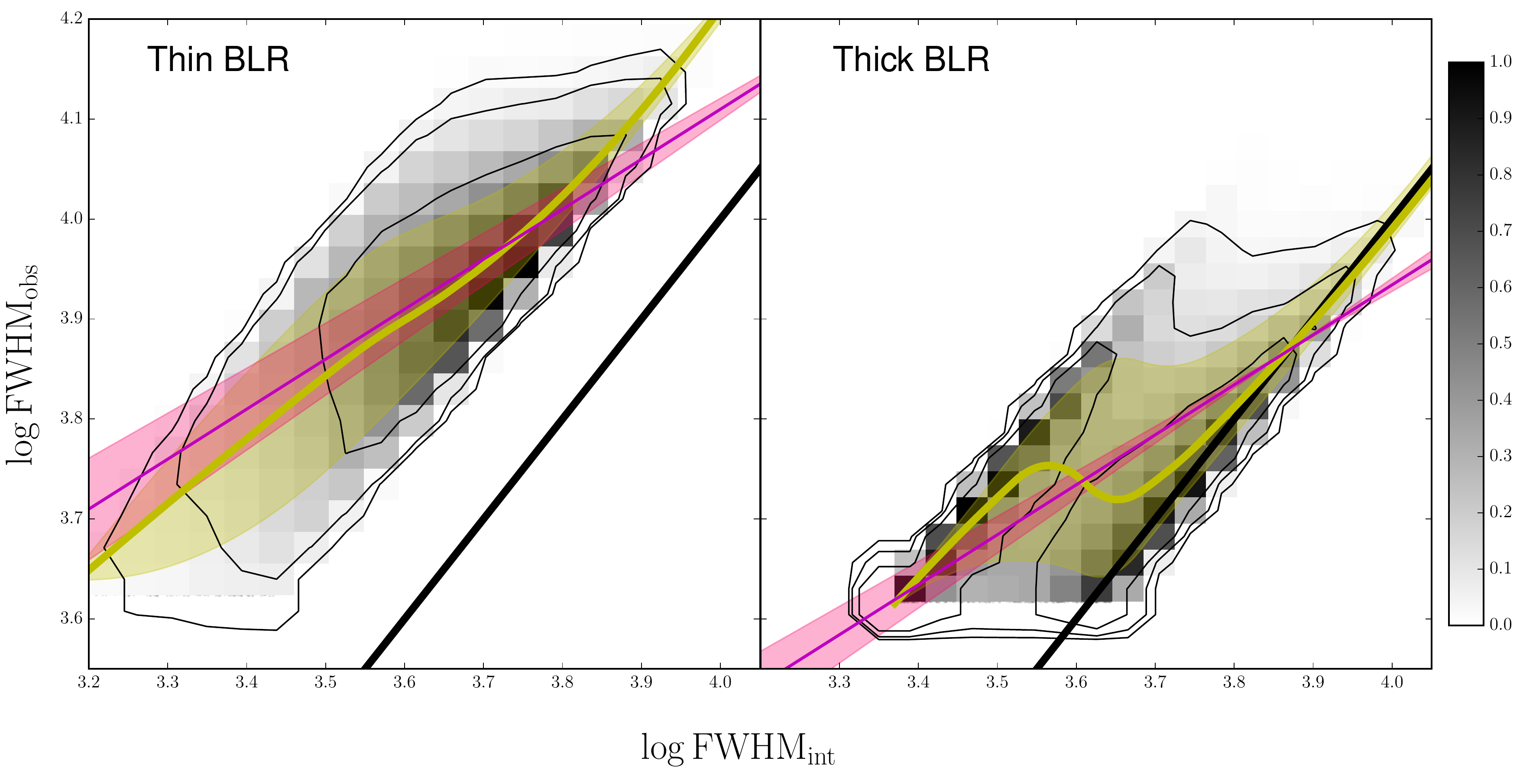}
  \caption{\textbf{\fwobs--\fwint\ bi--dimensional distributions for a thin and thick BLR.} 
   Bi-dimensional probability distribution of the intrinsic and observed \fwha\
  for a thin BLR (left) and a thick BLRs (right)  as predicted by the best-fit 
   models shown in Figure \ref{fig:E1}.  The darkest 
  regions show the most probable combinations of \fwint\ and \fwobs\
  values as quantified in the colour-bar. The thick black line is the 1:1 correlation.
  The thin black lines are the 68\%, 95\% and 99\% confidence limit contours centred
  around the maximum of the probability distribution.  The magenta line is the relation 
  $\fwint \propto \fwobshalf$ and the width of the magenta shadowed region accounts for 
  the  uncertainties in that relation. The solid yellow line is the 50\% regression 
  quantile of \fwint\ as a function of \fwobs\ for the theoretical probability density
  distribution and the yellow shaded region covers the 25\% to 75\% percentiles. We can 
  see that inclination closely reproduces the expected relation $\fwint \propto \fwobshalf$
  for the thin BLR but fails to reproduce it for the thick BLR case. }
   \label{fig:E3}
 \end{figure*}

  \begin{figure*}
  \renewcommand\thefigure{E5}
  \includegraphics[width=\textwidth,keepaspectratio]{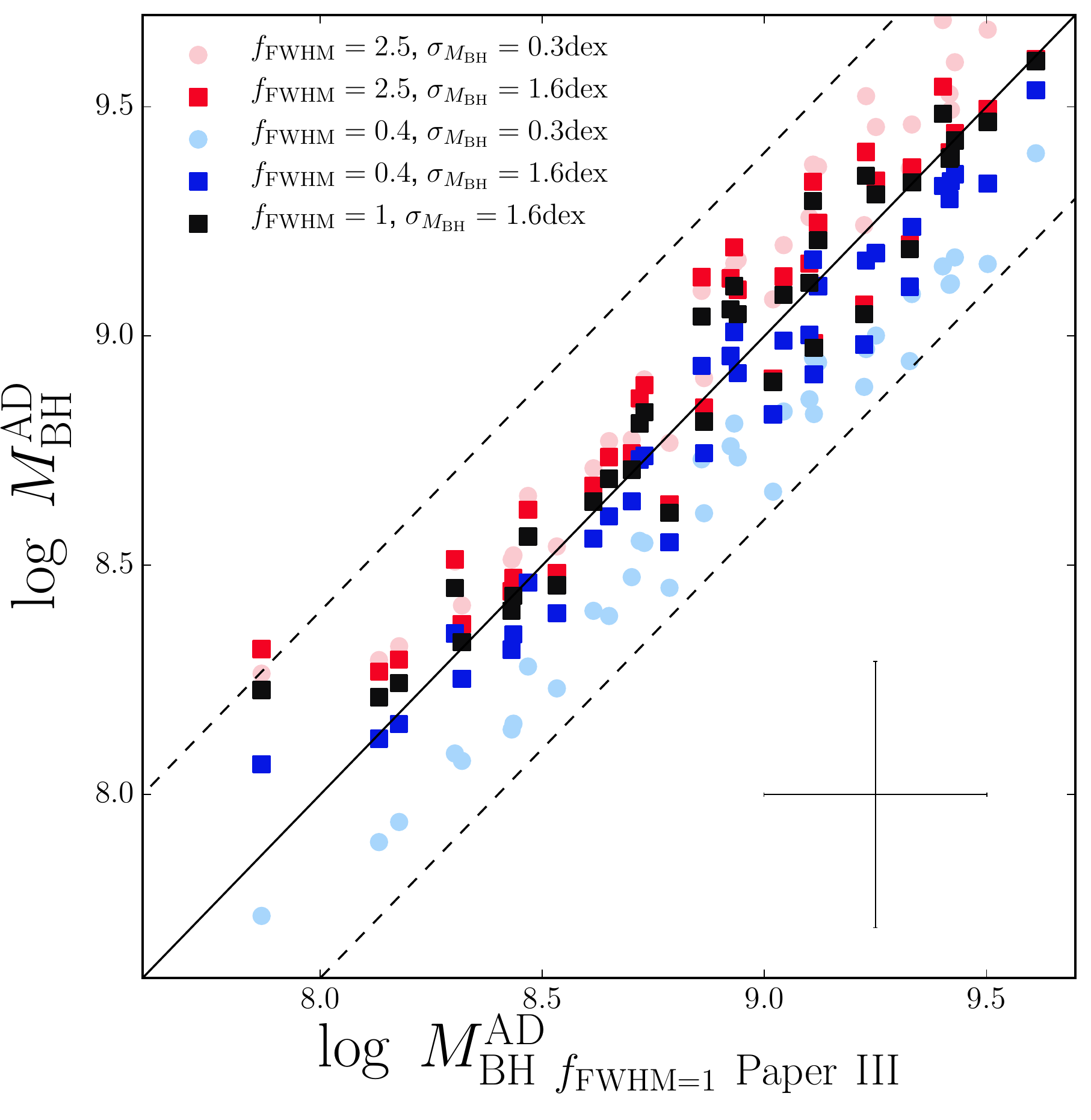}
  \caption{\textbf{Dependency of \MAD\ on the adopted values of $f$ and $\sigma_{\Mbh}$}.  Original \MAD\ values (from Paper III) vs
  Recalculated \MAD\ values obtained using different values for $f_{\fwhm}$ and the scatter in \Mbh\ ($\sigma_{\Mbh}$, as shown in the legend). The solid line represents the 1:1 relation  and the dashed lines represent $f=0.4$ and $f=2.5$.  As $\sigma_{\Mbh}$ increases \MAD\ values get closer to the 1:1 relation (with just one exception at the lowest mass). This indicates that $f=1$ is an appropriate initial choice. The cross symbol in the bottom right corner represents a  typical error bar in our \MAD\ estimations that accounts for the central 68\% of the marginalized posterior probability distribution. }
 \label{fig:E5}
 \end{figure*}

  \begin{figure*}
  \renewcommand\thefigure{E6}
  \includegraphics[width=\textwidth,keepaspectratio]{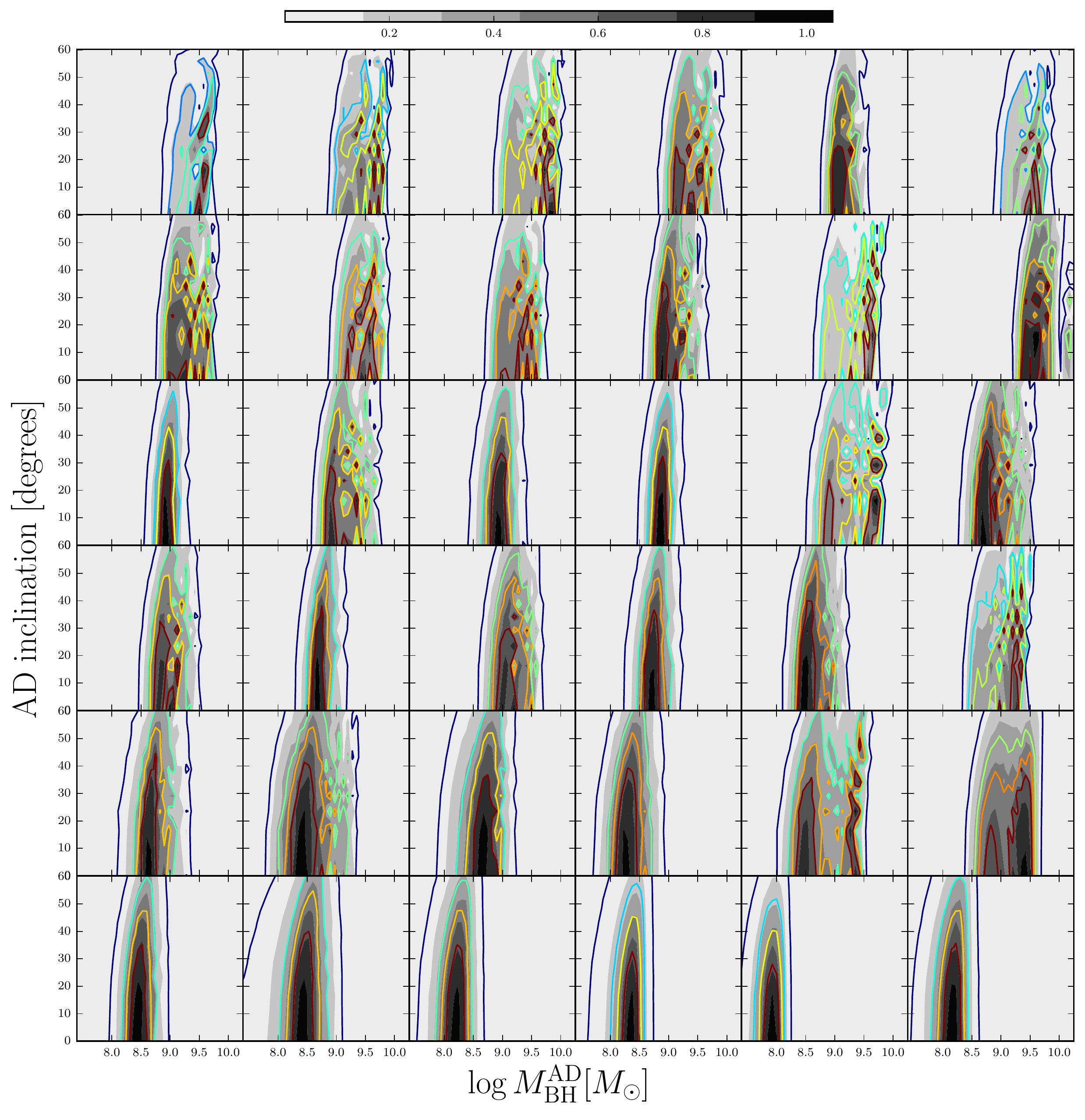}
  \caption{\textbf{\MAD\ versus AD inclination.}. Posterior bidimensional probability distribution of the accretion disc inclination versus the black hole mass in 36 objects of the sample.  Red, orange, cyan and blue lines represent the central 25, 50, 75 and 99 percentiles around the most probable point. The colour represents the relative probability normalized to the maximum probability for each object. }
 \label{fig:E6}
 \end{figure*}

  \begin{figure*}
  \renewcommand\thefigure{E7}
  \includegraphics[width=\textwidth,keepaspectratio]{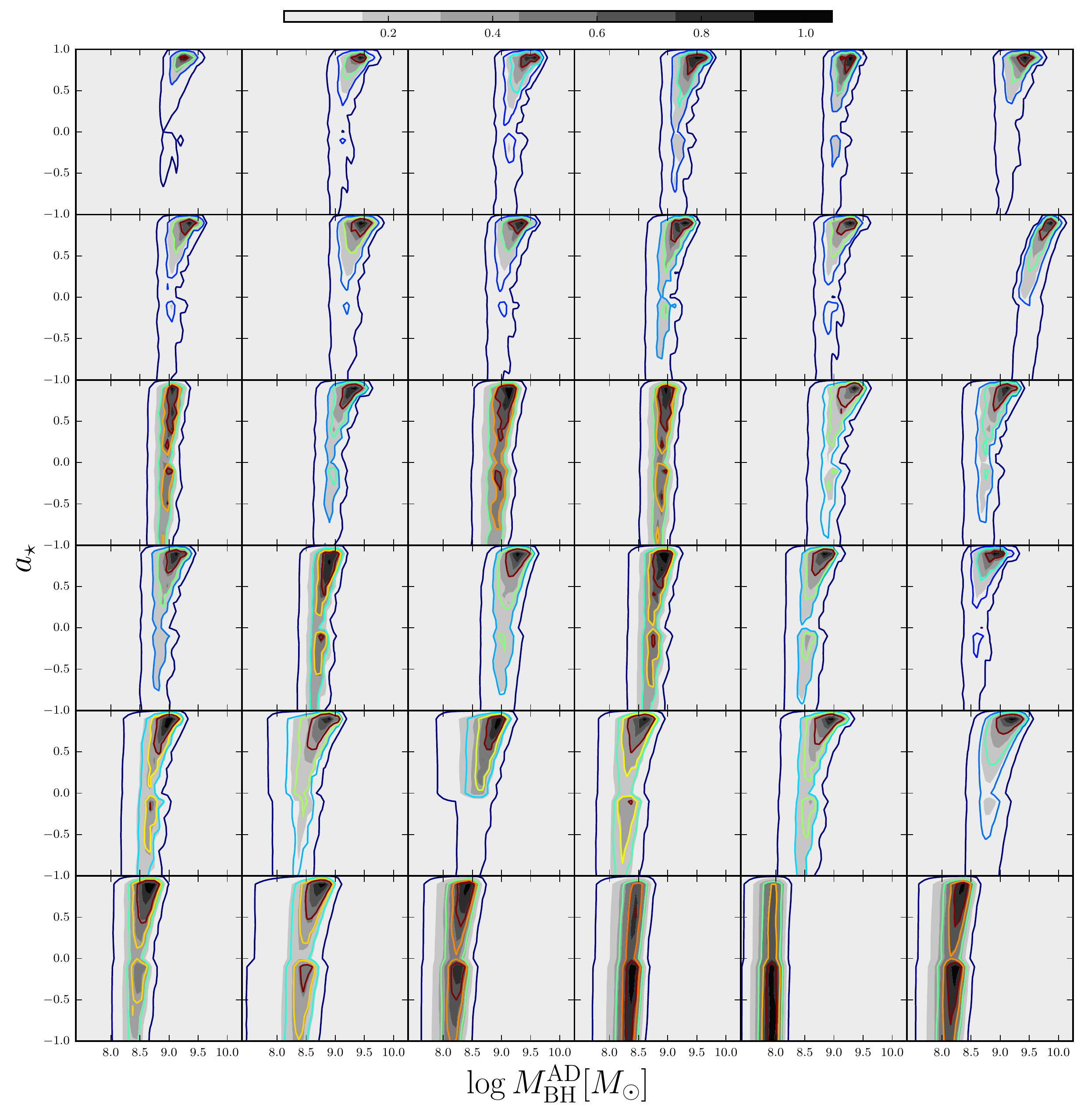}
  \caption{\textbf{\MAD\ versus the black hole spin}. Posterior bidimensional probability distribution of the spin (\astar) versus the black hole mass in 36 objects of the sample.  Red, orange, cyan and blue lines represent the central 25, 50, 75 and 99 percentiles around the most probable point. The color represents the relative probability normalized to the maximum probability for each object.   }
 \label{fig:E7}
 \end{figure*}

\clearpage

\begin{table*}
\renewcommand\thetable{E1}
\resizebox{1.0\textwidth}{!}{
\begin{tabular}{ccccccccccc}
\hline 
              & $\alpha_{\rm line}$ 
              & $\alpha_{\rm AD}$               & E                           & F                             &     $R^{2}$ & $M_{\rm BH}^{\rm AD}$ vs $M_{\rm BH}^{\rm SE}$ & $M_{\rm BH}^{\rm AD}$ vs $M_{\rm BH}^{\rm SE}(\rm corr)$ \\ 
              & [1] & [2] & [3] & [4] & [5] & [6] & [7] \\
              \hline 
$\rm H\alpha$ & 0.63     
& 0.66$\pm$0.03                  & 1.00$\pm$0.10               & 7.42$\pm$0.07               & 0.948   & 0.17                                               & 0.09\\ 
$\rm H\beta$  & 0.65         
& 0.69$\pm$0.04                  & 0.82$\pm$0.11               & 7.43$\pm$0.09               & 0.943   & 0.24                                               & 0.12\\
MgII          & 0.61       
& 0.68$\pm$0.05                  & 0.78$\pm$0.24               & 7.40$\pm$0.15               & 0.850   & 0.22                                               & 0.16\\
CIV           & 0.60         
& 0.61$\pm$0.06                  & 0.69$\pm$0.36               & 7.29$\pm$0.27                & 0.770   & 0.24                                              & 0.20\\ \hline            
\end{tabular}%
}
\caption{Properties of different correlations found for the emission lines of interest. Column [1]: power-law coefficient from reverberation mapping experiments (see 
Equation \ref{eqn:rblr}). Columns [2]-[4]: best fit parameters for linear regressions of the expression $\log \MADfl = \alpha_{\rm AD}\log\left( L_{\rm \lambda}\right)+
E \log {\rm FWHM\left({\rm line} \right)}+F$. Column [5]: $R^{2}$ values of the linear regressions. Columns [6] and [7]: scatter in the \MAD\ vs \MSE\ and \MAD\ vs \MSEcor\ 
correlations.}
\label{tab:MAD}
\end{table*}

\clearpage

\noindent \textbf{\huge Supplementary Information}

\section{Sample description}
\label{sec:data}
The sample we use in this letter consists of \Ntot\ type-I  AGN  selected
to be within a narrow redshift range around  $z\simeq1.55$.  For this sample
we  obtained high signal to noise ($S/N$) spectroscopic observations using
the VLT/X-Shooter spectrograph.  At the selected narrow redshift  range,
the X-Shooter spectrograph  covers a wide range from $\sim$1200 \AA\ to
$\sim$9200 \AA\ in the rest-frame.  The sample was selected to homogeneously
map the parameter space of \Mbh\  and $\lambda_{\rm Edd} =$ \LLedd\ within
the sampled region. The initial values of these quantities were obtained
from single-epoch (SE) calibrations\cite{TrakhtenbrotNetzer2012} of the
\Halpha\ broad emission line and its adjacent continuum. 
 
 The broad  spectral coverage and the high S/N in our sample allowed us to
 (1) re-calibrate, compare and test the performance of the different SE 
 \Mbh\ estimators using \Halpha, \Hbeta, \mgii\ and \civ \cite{MejiaRestrepo2016a};
 and (2)  model and confidently constrain the observed  Spectral Energy 
 Distributions (SEDs) in 37 out of 39 objects using  standard thin accretion
 disc models\cite{Capellupo2015,Capellupo2016}.  The output of the SED fitting provided
 alternative estimations for \Mbh, \Mdot , $\lambda_{\rm Edd}$  and a realistic
 estimate of  \astar. For the sake of simplicity, hereafter when referring to paper 
 I, II and III we will be citing references 23, 26 and 24, respectively.

\section{Estimating \Mbh }
\label{sec:Mbh}

 In this section we briefly describe the two alternative approaches
 that we followed to derive \Mbh\ and comment on the sources of uncertainties 
 of each method.

\subsection{Single Epoch \Mbh\ estimates}
\label{subsec:virial}
We used the black hole masses obtained in paper II from the new calibrations of the single epoch (SE)
black hole mass estimators for  the broad $H\alpha$, $H\beta$, $\rm MgII$ and 
$\rm CIV$ emission lines. In particular, we used the coefficients 
of the first two columns of Table 7 in Paper II.  The underlying assumption in SE estimations
is that Equation\ref{eqn:virial_eqn} holds for all broad emission lines and   
$V_{\rm BLR}$ can be estimated from the \fwobs\ of the line in question using Equation
\ref{eqn:virial_eqn}. We used $f=1$ as suggested from \Mbh-Stellar dispersion 
calibrations\cite{Woo2015}. $R_{\rm BLR}$ is obtained from the  calibration of the 
$R_{\rm BLR} - L$ relation obtained from various RM 
studies\cite{Kaspi2000,Kaspi2005,Bentz2009,Bentz2013} which can be written as:

\begin{equation}
R_{\rm BLR} = R_{\rm BLR}^{0}  \left(\frac{L_{\lambda}}{10^{44}\ergs}\right)^{\alpha_{\rm line}}
\label{eqn:rblr}
\end{equation} 
where, $R_{\rm BLR}^{0}$ is the normalization constant which for the case of the \Hbeta\ 
line, and for $\lambda=5100$\AA,  is 538 light-days\cite{MejiaRestrepo2016a}.

As we  briefly discussed in the letter, the simple SE mass determination method
is limited in various important ways:

\begin{enumerate}

 \item The $R_{\rm BLR}-L$ relation has been obtained from a relatively small sample
 of low-z ($z \lesssim 0.3$) Seyfert I galaxies and low luminosity quasars ($\Lop \lesssim 10^{46}\ergs$,
 where $\Lop \equiv 5100\text{\AA} \times L\left(5100 \text{\AA}\right) $). Therefore,
 extrapolation of the $R_{\rm BLR}-L_{5100}$ relation is needed  to estimate \Mbh\ in
 high luminosity objects at high-z. Moreover, the intrinsic scatter in the 
 $R_{\rm BLR}-\Lop$ relation is affected by intrinsic luminosity variations 
 as well as by the disc inclination to the line-of-sight\cite{DavisLaor2011,NetzerTrakhtenbrot2014,Capellupo2015}.

 \item The re-calibration of the \Hbeta-based single epoch method to other broad
 emission lines like \Halpha, \MgII\ and \CIV\ induces intrinsic dispersion that
 can be as high as 0.5 dex  for the \CIV\ line\cite{MejiaRestrepo2016a}.
\item The dependence of $f$ on inclination is a major source of uncertainty.
This has been  explored in numerous papers. A recent paper used a sample of  about
600 local SDSS type1-AGN to compare the \Mbh\ estimations  derived from the
\Mbh-stellar dispersion relation (\MSIGMA) with those derived from the single
epoch method\cite{ShenHo2014}. They found that 
$f_{\sigma^{\star}} \equiv \MSIGMA/(G^{-1} \RBLR \fwhb^{2})$  is anti-correlated 
with \fwhb, and argued that this is a manifestation of the line-of-sight 
inclination in a flat, disc-like BLR.  Earlier works also suggested an anti-correlation
between the radio loudness of sources and the observed \fwhb\cite{Wills1986,Runnoe2014}.
Assuming that radio jets in AGN are aligned with the axis of symmetry 
of the BLR  and that the flat BLR is aligned with the disc, their results strongly suggests 
that the BLR in radio-loud AGN  are considerably flattened.

 \item There are questions regarding the validity of virial equilibrium 
 of the BLR  material. Earlier results about AGN with multiple emission
 measurements  (i.e., NGC3783, NGC5548,  NGC7469 and 3C390.3) show that the
 velocity radial profiles that are in good agreement with the
 expectations for a Keplerian system\cite{PetersonWandel2000, OnkenPerterson2002}
 (i.e., $V_{\text{BLR}}(r)\propto r^{-1/2}$). Additionally, in some 
 velocity resolved RM experiments, the blue wing of the \Hbeta\ line
 has been observed to lag behind the red wing, which generally rules
 out significant outflow of both high- and low-ionization lines\cite{DoneKrolik1996, UlrichHorne1996, Sergeev1999}. 
 However, more recent RM observations revealed diverse kinematics of the BLR  
 including  inflows, outflows and virialized gas\cite{Denney2009, Denney2010}. 
 
\item The use of a single value of $f$ for measuring \Mbh\ in sources 
that are not part of  RM samples introduces an additional uncertainty which 
results from the fact that the \fwobs\ measured from single-epoch  spectra
are systematically larger than those measured from the RMS profiles during a RM campaign\cite{Collin2006}.
This can be easily verified by comparing the RMS \fwhb\ \cite{BentzKatz2015}
with the one measured from the mean spectrum of the same sources\cite{Du2015}. 
This is also true when the standard deviation of the lines ($\sigma_{\rm obs}$) is used 
instead of the \fwobs\ and has not been taken into account, properly, in many studies. 
For example, from the results published in a recent paper\cite{Batiste2017} we 
obtain $\fwhm_{\rm mean}/\fwhm_{\rm rms}=1.17^{+0.37}_{0.15}$.

\item The line shape parameter $\fwobs/\sigma_{\rm obs}$  provides information on the 
structure and  kinematics of the BLR. For instance, $\fwobs/\sigma_{\rm obs} \sim 3.4$ is found 
for a spherical shell of clouds moving with fix a velocity and random orientations,  
$\fwobs/\sigma_{\rm obs} \sim 3$ is found for an face-on rotating ring with fixed velocity,
$\fwobs/\sigma_{\rm obs}=2.35$ corresponds to Gaussian profiles, $\fwobs/\sigma_{\rm obs}  
\sim 2$ is found for a face on rotating Keplerian disc, $\fwobs/\sigma_{\rm obs}\sim 1$ ($\to 0$) 
corresponds to Lorentzian profiles due, for example, to turbulent motions, as well as logarithmic profiles 
which can be caused by in-/outflow motions, and $\fwobs/\sigma_{\rm obs}\sim 0.98$ corresponds 
to exponential profiles caused by electron scattering within the photo-ionized BLR gas\cite{Kollatschny2013}.
$\fwobs/\sigma_{\rm obs}$ is known to vary within a significant range\cite{Collin2006, Kollatschny2011,Peterson2004} 
suggesting  that BLR profiles are not universal and that the virial factor is far from being a constant value. 

\end{enumerate}

\subsection{Black Hole Mass Estimates from SED fitting}
\label{subsec:AD}

As previously mentioned, in papers I and III we recently implemented an alternative method to
estimate the black hole mass in type1-AGN based on the fitting of the SED 
of the accretion discs, using a geometrically thin, optically thick accretion
disc model, and obtained successful fits in 37 out of 39 objects in our sample. 
 The model is  fully determined by \MAD, 
\astar, $\dot{M}$, the AD inclination with respect  to the line-of-sight 
($i_{\rm AD-LOS}$), and intrinsic $A_{\rm V}$. The procedure consisted 
of a Bayesian  minimization over a grid of models covering a range in values for
these parameters.  We assumed Gaussian priors for \MSEha\ and $\dot{M}_{\rm SE}$. Means 
were given by the single-epoch estimations for each object and standard deviations of 
0.3 and 0.2 dex were adopted, respectively. \MSEha\ and $\dot{M}_{\rm SE}$ were calculated  
assuming a virial factor $f_{\fwobs}=1$. Flat priors were assumed for the remaining 
model parameters. The role of the priors is to penalize models which deviate significantly 
from the observational estimations of $M_{\rm BH}\left({\rm H}\alpha \right)$ and 
$\dot{M}_{\rm SE}$, but allow a symmetric parameter search on either side of the means.
In the analysis presented in this letter  we use the results obtained in paper III. To completely avoid complications in SED fittings resulting from unknown opacity and other effects in the disc atmosphere, for our Bayesian fitting we used only the X-shooter spectra with $\lambda>1200$\AA. At such wavelengths, our models are adequate and spectral features that may influence the spin parameter determination are not included in the fit.

We have investigated the possibility that our initial choices of $f_{\fwobs}$ and the standard 
deviation in \MSEha\ ($\sigma_{\Mbh}$), affect the resulting \MAD. As shown in Figure \ref{fig:E5},
this is not the case. In this Figure, we compare the resulting \MAD\ for a large range of initial guesses
in $f$ and $\sigma_{\Mbh}$ with the original \MAD\ values obtained for our sample in paper III 
(where $f_{\fwobs}=1$ and $\sigma_{\Mbh}=0.3$ dex were assumed). We find that for large enough
$\sigma_{\Mbh}$ ($\gtrsim 0.8$ dex) there is basically no difference between the  \MAD\ values 
obtained with our initial choices and those \MAD\ values obtained with an initial $f$ varying 
over a large range (0.4-2.5). Motivated on the previous findings, we re-ran our code using completely
random mass priors within the $\log \Mbh$  interval of [8,10] and $\sigma_{\Mbh}=1.6$ dex.  We find that
(1) the resulting ``random-AD masses"  do not show significant correlation with the assumed random priors
($r_{\rm s}=0.24$, $P_{\rm s}=0.14$) and, crucially, that (2) random-AD masses  are  significantly correlated
and consistent with SE masses from paper II with $r_{\rm s}=0.69$ and $P_{\rm s}=2\times10^{-6}$.  All these
tests confirm the robustness of the BH mass measurements obtained by fitting AD SEDs as well as their comparable
accuracy and good agreement and with SE mass measurements.

We also tested the reliability of the \fad\ anti-correlation with \fwobs\ found using our 
Bayesian algorithm. We explored whether the assumption of flat priors for 
$i_{\rm AD-LOS}$ and $A_{\rm V}$ had an impact on our results. We tested various
Gaussian priors on $\cos i_{\rm AD-LOS}$ assuming as central values some randomly 
assigned numbers and  different intrinsic scatters of 0.1, 0.2 and 0.3 dex. We 
also assumed Gaussian priors on $A_{\rm V}$. The central values were obtained
from the recent calibrations of $A_{\rm V}$  based on the $L(\Halpha)/L(\Hbeta)$
ratio\cite{Baron2016}. The intrinsic scatter were varied from 0.1 to 0.3 dex. 
In all cases we  recover the anti-correlation between $ \log f_{\rm AD}$ and
the \fwobs\ for the Balmer lines with similar statistical significance. We also
used \MSEciv\ as the central value for the \Mbh\ prior and tested using the
median values, instead of the mean, for \Mbh\ and \Mdot.  In both cases we 
recover the aforementioned anti-correlation with similar statistical
significance. We conclude that our findings are not an artefact of the fitting code.

One important drawback from our modelling is a large degeneracy between  the 
accretion rate and the inclination angle of the disc. 
For a given flux, larger inclinations will return larger intrinsic luminosities 
which in turn will return larger accretion rates. Fortunately, the derived black 
hole mass does not strongly depend on either inclination nor accretion rate
and the mass estimates are consistent within 0.1 dex regardless of the final 
derived inclinations and accretion rates. This is confirmed in Figure \ref{fig:E6} that shows  the posterior bi-dimensional probability distribution of the accretion disk inclination versus \Mbh\ in 36 out of 37 objects of our sample. Although the inclination is mostly unconstrained, the derived black hole mass is practically independent of the assumed inclination. As a consequence, the derived disc inclinations are very uncertain and are {\it not} good indicators of the real
inclinations of the disc and consequently of the  flat BLRs. Therefore, these values are not used as proxies for the inclination of the BLR and the determination of the virial factor.

Finally, our results also show that the derived \Mbh\ is mostly independent of \astar. This is confirmed in Figure \ref{fig:E7}
that shows  the posterior bi-dimensional probability distribution of \astar\ versus \Mbh\ in 36 out of 37 objects of our sample. It can be observed that although the spin is not tightly constrained, the black hole mass is restricted to a narrow range. This indicates that our AD mass determinations are not degenerated with \astar. The reason for this is that the spin affects mostly the ionizing UV continuum emission while the mass depends mainly on the optical part of the AD spectrum.

\section{$f$ as a function of line width}

In Table \ref{tab:t1} we present the correlation coefficients of the \fad--\fwobs\ 
and \fad--$R_{\text{BLR}}\fwobstwo/G$ correlations for all the broad emission 
lines considered here. In all cases we find that the correlations associated with
the \fwobs\ are stronger than those associated with $R_{\text{BLR}}\fwobstwo/G$.
This suggests that the \fwobs\ correlations are not inherited from the definition
of \fad. In order to prove this we conducted the Williams's Test\cite{DunnClark1969}. 
Given a sample size, this test computes the statistical significance of the 
difference between the correlation coefficients of two correlations 
that have one variable in common. In this case the two dependent correlations are 
\fad--\fwobs\ and \fad--$R_{\text{BLR}}\fwobstwo/G$, while the common variable is \fad. 
Our results indicate very different correlation coefficients for the \fad--\fwobs\
and the \fad--$R_{\text{BLR}}\fwobstwo/G$ correlations (see Table \ref{tab:t1}), with a 5-$\sigma$ significance for the
\Halpha\ line, 4-$\sigma$ significance for the \Hbeta\ and the \MgII\ lines and a 3-$\sigma$
significance for the \CIV\ line. This confirms that the correlations associated
with the \fwobs\ of the broad emission lines are indeed much stronger than those
that by definition are associated with $R_{\text{BLR}}\fwobstwo/G$. 

Single epoch \Mbh\ can also be estimated using $\sigma_{\rm obs}$ instead of \fwobs. 
Obviously, in that case the virial factor has a different numerical value since \fwhm\ 
can be significantly different from $\sigma_{\rm obs}$ (e.g., for a Gaussian line profile 
$\fwobs=2.35\ \sigma_{\rm obs}$). We tested whether the associated 
$f_{\rm AD}\left( \sigma_{\rm obs}\right) \equiv M_{\rm BH}^{\rm AD}\ 
/\left( G^{ - 1} R_{\rm BLR}\ \sigma_{\rm obs}^2\right)$ is also anti-correlated with 
$\sigma_{\rm obs}$ and confirmed statically significant anti-correlations using all 
four emission lines. However, in this case there is no statistical difference between 
the \fad--$\sigma_{\rm obs}$ and \fad--$R_{\text{BLR}}\sigma_{\rm obs}^{2}/G$ correlations. 
This is most likely due to the larger uncertainties associated with the measurement of 
$\sigma_{\rm obs}$ in our sample\cite{MejiaRestrepo2016a}.

To determine how \MAD\ depends on the \fwobs\ of the lines and the associated
$L_{\lambda}$ used in single epoch mass determinations methods, we used
the following expression:
{
\small
\begin{eqnarray}
\log \MADfl \equiv  \alpha_{\rm AD}\log\left( L_{\rm \lambda}\right)+E \log {\rm  FWHM\left({\rm line} \right)}+F
\label{eqn:MADfl}
\end{eqnarray}
}
and implemented an ordinary bi-variate least square linear regression to determine 
the coefficients in the equation. We summarize the results in Table \ref{tab:MAD},
where we also show $\alpha_{\rm line}$, which represents the slope of the power-law 
coefficient of $L_{\rm \lambda}$ in Equation \ref{eqn:rblr}. We also list the scatter
between \MAD\ and \MSE\ as well as the scatter between \MAD\  and the {\it corrected\/}
\MSE\ ($\MSEcor \equiv \MADfl$) after the dependency of \fad\ on \fwobs\ is taken into account.  
In the case of the Balmer lines, the  scatter is reduced by about a factor 2. 
Thus, correcting for the correlation between $\log f_{\rm AD}$ and the \fwobs\ of the 
Balmer lines provides an important improvement in our \Mbh\ estimations.  
 
The results of the linear regressions presented in Table \ref{tab:MAD} highlight
two important findings. First, $\alpha_{\rm AD}$ and $\alpha_{\rm line}$ are basically
indistinguishable from each other. This indicates that $L_{\rm \lambda}$ has almost no
impact in the deviation of \MSE\ from \MAD\ and that \MAD\ preserves its dependency on 
$R_{\rm BLR}$. Second, the dependence of \MAD\ on the observed \fwobs\ of the Balmer
lines is close to linear rather than quadratic, as expected from the virial relation.

\section{Inclination as the source of the $f$--\fwobs\ correlation }

In this section we present different tests that we carried out to determine whether 
inclination is driving the correlation between $f$ and \fwobs.

Hereafter when referring to $ \log f_{\rm AD}$, \MSE\ and  \fwobs\ we mean
$\log f_{\rm AD} \left({\rm \Halpha}\right)$, $M_{\rm BH}^{\rm SE}\left(\fwha \right)$
and  the observed value of \fwha, unless otherwise specified. The reason to select
the \Halpha\ line instead of the \Hbeta\ line for the following analysis is the better 
S/N and hence more accurate measurements of \fwha\ in our sample. As shown in earlier
works, \fwobs\ in both Balmer lines are the same within uncertainties
\cite{GreeneHo2005,MejiaRestrepo2016a}.

The anti-correlation between $\log f_{\rm AD}$ and \fwobs\ could be explained by 
the inclination of the axis of symmetry of a disc-like 
BLR with respect to the line-of-sight (LOS). If we consider the median LOS inclination,
$i_{\rm median}$, at which Type-1 AGN are typically observed, we can also define 
a median virial factor $f_{\rm median}$ at which the SE \Mbh\ calibration represents 
an accurate black hole mass for objects observed at $i_{\rm median}$. Objects with 
narrower than usual broad emission lines are more likely observed at $i<i_{\rm median}$ 
(face-on orientations) and objects with broader than usual emission are more likely 
observed at $i>i_{\rm median}$ (edge-on orientations). This will produce too large 
(too small) SE mass estimates for objects with very broad (very narrow) emission lines,
and would translate into a virial factor that anti-correlates with the line FWHMs. 

The inclination hypothesis is also consistent with recent work that found that 
$f_{\sigma^{\star}}\equiv\MSIGMA/(G^{-1} \RBLR \fwhb^{2})$ is anti-correlated with
\fwhb\cite{ShenHo2014}. Here, \MSIGMA\ is the black hole mass obtained from the 
correlation between \Mbh\ and the stellar dispersion of the spheroidal component 
in galaxies. Analogously, an earlier work compared the virial black hole masses 
with black hole mass estimations obtained from the relation between black hole 
mass and the luminosity of the host-galaxy spheroidal component\cite{Decarli2008b}.
Their results also show a clear anti-correlation between $f$ and the \fwobs\ of the 
broad emission lines that is interpreted by the authors as a BLR line-of-sight
inclination bias.

There is further evidence that favours the hypothesis that LOS inclination is
biasing SE \Mbh\ estimations. As already pointed out, previous works found 
that the \fwhb\ is significantly anti-correlated with radio core dominance in
radio-loud quasars\cite{Wills1986,Runnoe2014}. This is consistent with \Hbeta\
emitting gas in a flattened configuration. In this scenario core-dominated objects
(with their radio emission being Doppler-boosted along the LOS) correspond to 
face-on discs that typically show narrow \Hbeta\ profiles, while lobe-dominated
objects (lacking Doppler-boosting) correspond to edge-on discs, that typically
show broad \Hbeta\ profiles. In this scenario, the BLR is flat and the general 
plane of motion is similar to the plane of rotation of the central disc. In 
addition to this, there is accumulated evidence 
in the literature favouring a disc-like geometry for the 
BLR\cite{MclureDunlop2001,MclureDunlop2002,Laor2006,Decarli2008a,Pancoast2014}.

For a disc-like BLR with a thickness ratio $H/R$ and inclination $i$ with 
respect to the line-of-sight we will have $\fwobs = \fwint \times
\sqrt{\sin^{2}(i)+(H/R)^{2}}$. Thus, for an ensemble of 
randomly orientated BLRs  the final distribution of \fwobs\ will depend on (1) 
the intrinsic \fwint\ distribution and (2) the range of possible random 
orientations at which the BLR can be observed, both of which are, a priori, 
not known.

To check the inclination hypothesis we first need to determine the distribution 
of \fwint\ that is consistent with the probability density distribution (PDF) 
of the observed \fwobs. We then need to test whether it is possible to recover 
the anti-correlation of $f$ with \fwobs\ and the linear dependence of \Mbh\ on 
${\rm FWHM}_{\rm obs}$, as derived in this letter. In other words, we need to 
test whether a population of randomly generated inclinations and \fwint\ that 
satisfy the PDF of \fwobs, can also account for: 
\begin{equation}
f \propto \fwobsminus
\label{eqn:fFWHM}
\end{equation}
and at the same time:
\begin{equation}
\fwint \propto \fwobshalf
\label{eqn:MADobs}
\end{equation}

It is important to note that both predictions should be satisfied to guarantee 
inclination as the driving mechanism of the observed correlation between $f$ and 
\fwobs. The reason for this is that each of these expressions tests the dependency 
between \fwobs\ and the two independent distributions determined to reproduce \fwobs:
$\sin(i)$ and \fwint. While Equation (5) tests the dependency between \fwobs\ and 
$\sin^{-2}(i)$ (which is proportional to $f$), Equation (6) tests the dependency 
between \fwobs\ and \fwint.

We first assumed a thin BLR by taking $H/R=0$. We computed the PDF as the product
of two independent random variables\cite{Glen2004} and applied it to the special 
case where ${\rm FWHM}_{\rm obs}={\rm FWHM}_{\rm int} \times \sin\left(i\right)$\cite{LopezJenkins2012}. 
For the \fwint\ distribution, we assumed an underlying truncated normal 
distribution with certain mean  (FWHM$_{\rm mean}$) and dispersion 
(FWHM$_{\rm std}$). Our normal distribution was truncated to allow \fwint\ 
to vary between 1000 and 30000 \kms. We also assumed that our sample is 
limited to objects with line-of-sight inclination angles between $i_{\rm min} = 
0^{\circ}$ and $i_{\rm max} = 70^{\circ}$, with $i_{\rm max}$ determined by the torus 
opening angle. For an optimal exploration of the parameter space we ran a Monte Carlo Markov 
Chain simulation using the python code EMCEE\cite{emcee2013}. For the simulation
we used 20 independent walkers and 5000 iterations that mapped a total of
$10^{5}$ models.

In the left panel of Figure \ref{fig:E1} we compare the observed cumulative
${\rm PDF}\left({\rm FWHM}_{obs} \right)$ and its uncertainty (magenta thin line 
and shadowed region, respectively) with the predicted cumulative PDF from the model
with the highest posterior probability (black line). The parameters of this 
model are: $i_{\rm min}=19^{\circ}$, $i_{\rm max}=45^{\circ}$, ${\rm FWHM}_{\rm mean}=8500$,
${\rm FWHM}_{\rm std}=2150$, ${\rm FWHM}_{\rm min}=4200$ and 
${\rm FWHM}_{\rm max}=30000$.  Our model successfully reproduces the observed
cumulative PDF. However, a simple normal distribution (red dashed line) is 
also consistent with the data and cannot be rejected. We also determined the best 
fit model for a distribution with FWHM$_{\rm std} = 0$, i.e., effectively a
single velocity. This model (dashed blue-line) is able to reproduce the 
distribution at low values of \fwobs, but it is unable to account for 
the distribution at large velocity widths.

First, we tested whether our thin BLR model is successful in reproducing the $f$--\fwobs\ 
distribution seen in the data (i.e., Equation 5). In the left panel of Figure \ref{fig:E3}
we show the predicted bi-dimensional probability density distribution of the virial 
factor and the observed \fwha\ as predicted by the thin BLR model. The Figure includes
contours showing 25\%, 50\%, 75\% and 99\% confidence limits contours (black-thin lines)
centred around the maximum probability point. We also superposed the data from 
in Figure \ref{fig:f2} (open-blue circles). The magenta line represents the derived
relation $f = \left(\fwha/4000\ \kms \right)$. The thick yellow line is the median of the 
$f$--\fwobs\ distributions derived using a quantile non-parametric spline regression\cite{NgMaechler2007}.  
Analogously, the blue-dashed lines represent the 25\%, 50\% and 75\% quantiles of the 
observational distribution. To obtained these quantiles, for each observed data 
we randomly generated 1000 points following the error distributions in \fad\ and \fwha\ and then 
applied the COBS method to characterize the resulting distribution. We can notice that the median 
(50\%-quantile) of the theoretical and observational distributions are in very good agreement.
The scattered open-blue circles also show excellent agreement with the the bi-dimensional 
probability density function from the best model. Explicitly, we find that from our 37 objects,
21\% fall inside the central 25\% confidence level region, 51\% fall inside the 50\% confidence 
level region, 74\% fall inside the 75\% confidence level region, and 85\% fall inside the 99\% 
confidence level region.

Next, we tested for the same thin BLR model whether it is possible to recover the predicted
relation between \fwobs\ and  \fwint\ (i.e., Equation 6). In the left panel of Figure \ref{fig:E3} 
we show the predicted bi-dimensional probability distribution of \fwint\ versus \fwobs\ using the 
model with the highest posterior probability. The magenta solid line and magenta shadowed region represent
the expected \fwint\ $\propto$ \fwobshalf\ relation and 1-$\sigma$ uncertainties, respectively. The 
solid-yellow line and yellow shadowed region represent the median \fwint--\fwobs\ and errors from 
the simulated bi-dimensional distribution. A good agreement is found between the simulations and the 
predicted relations. This implies that we are also able to recover the relation $\fwint \propto 
\fwobshalf$ for the thin BLR model.

In order to test the effects introduced by a thick BLR ($0<H/R<1$), we assumed a single $H/R$ for all 
objects and followed the same steps outlined for the case of a thin BLR. We found that a wide range in BLR 
thickness ratios ($H/R<0.5$) is able to reproduce the cumulative \fwobs\ PDF. However, objects with large 
thickness ratios clearly fail to reproduce the bi-dimensional distributions of $f$--\fwobs\ and 
\fwint--\fwobs, as can be seen in the right panels of Figures \ref{fig:E2} and \ref{fig:E3}. We 
generally find that only relatively thin BLRs, i.e., those with $H/R<0.1$, are able to reproduce 
both bi-dimensional distributions and the cumulative \fwha\ PDF. In particular, for a BLR with 
$H/R\to 0$, we find that the derived $\fad$ values constrain the range of inclinations at which the BLR 
is observed in our sample to $15^{\circ} \lesssim i \lesssim 50^{\circ}$. This upper limit is consistent 
with typical expectations of a central torus hiding the BLR. We also find that the median virial factor 
in our sample, $f=0.95$, corresponds to a median orientation of $i_{\rm median}=31^{\circ}$.

In summary, our results show that a population of randomly orientated, thin BLRs can successfully
reproduce our observations.  We can thus conclude that inclination is very likely the main reason for 
the observed $f$--\fwobs\ correlations.

\section{Radiation pressure effects}

We finally considered the possibility that non-virial BLR motions or
winds induced by radiation pressure force might cause the observed \fad-\fwobs\
dependency. We first tested a simple model that assumes that the BLR is 
composed of homogeneous clouds that are optically thick to ionizing radiation 
but optically thin to electron scattering. The model predicts a dependency 
between the virial factor and the the normalized accretion rate, $\lambda_{\rm Edd}$, 
of the form: $f = f_{1} \left[1+K\ \lambda_{\rm Edd}\right]$, where $f_1$ is the true 
virial factor and $K$ depends on the fraction of ionizing radiation and the column density 
of the gas clouds that is assumed constant along the entire BLR\cite{Marconi2008}. 
From this expression we can see that \MSEE\ underestimates the actual \Mbh\ 
as $\lambda_{\rm Edd}$ increases. Equivalently, \fad\ should increase as 
$\lambda_{\rm Edd}$ increases. However, we find no clear correlation between 
$\lambda_{\rm Edd}$ and \fad\ in our data ($r_{\rm s}=0.2$, $P=0.23$),
and therefore radiation pressure effects, as prescribed by this model, are not
present in our objects. Note however that our sample is restricted to a 
relatively small range of $\lambda_{\rm Edd}$ (from $\lambda_{\rm Edd}=0.01$ 
to $\lambda_{\rm Edd}=0.3$, corresponding to a variation by a factor of 30). 

A more recent model considers the effects of radiation pressure in a more 
realistic BLR composed of pressure confined clouds, hence allowing the gas 
density of individual clouds to decrease with distance to the central black 
hole\cite{NetzerMarziani2010}. In this model the system is still bound by
gravity and \fwobs\ becomes smaller with increasing $\lambda_{\rm Edd}$. The 
reason for this trend is that as $\lambda_{\rm Edd}$ increases, the clouds spend
more time at large distances from the black hole, therefore increasing the median 
$R_{\rm BLR}$ and decreasing the median BLR Keplerian velocities. To account for 
this effect, the authors of this model proposed a modified expression for $R_{\rm BLR}$:
\begin{equation}
R_{\rm BLR} = R_{\rm BLR}^{0}\left[  a_{1}L_{\lambda}^{\alpha_{\rm line}} +a_{2}\left(L_{\lambda}/M_{\rm BH}\right)\right]
\label{eqn:rblro}
\end{equation} 
where $a_{1}$ and $a_{2}$ are constants. The first term accounts for the observational 
relation described in Equation \ref{eqn:rblr} and the second term represents a radiation 
pressure perturbation quantified by $L_{\lambda}/M_{\rm BH}\propto \lambda_{\rm Edd}$. 
When replaced into the virial mass equation (Equation \ref{eqn:virial_eqn}) this relation
leads to a simple quadratic equation on \Mbh\ with solution: 
\begin{equation}
\Mbh^{\rm rad}=\frac{a_{10}}{2}L^{\alpha_{\rm line}}\fwobs^{2}\left[1+\sqrt{1+\frac{4\ a_{20} L_{\lambda}^{1-2\alpha_{\rm line}}}{a_{10}^{2}\ \fwobstwo}}\right]
\label{eqn:rad_mass}
\end{equation}
or equivalently:
\begin{equation}
f_{\rm rad}\propto\left[1+\sqrt{1+\frac{4\ a_{20} L_{\lambda}^{1-2\alpha_{\rm line}}}{a_{10}^{2}\ \fwobstwo}}\right]
\label{eqn:rad_factor}
\end{equation}
where \Mbh$^{\rm rad}$ and $f_{\rm rad}$ are the black hole mass and virial factor for a 
radiation  pressure dominated BLR. $a_{10} = a_{1} f_{0} \ R_{\rm BLR}^{0} G^{-1}$, 
$a_{20} = a_{2} f_{0} \ R_{\rm BLR}^{0} G^{-1}$, and $f_{0}$ is a normalization constant. 
In the case when $4\ a_{20} L_{\lambda}^{1-2\alpha_{\rm line}}/a_{10}^{2}\ \fwobstwo \gg 1$ 
this would result in a close agreement with the inverse proportionality between \fad\ and
\fwobs\ found in our data. Given that $\alpha_{\rm line}$ is found to be $~\sim 0.6$ for 
all lines (Table \ref{tab:MAD}), this would translate into an explicit dependency of 
$f$ on $L_{\lambda}$.  We would then expect that the scatter in the \fad--\fwobs\ relation 
should be driven by $L_{\lambda}$. In Figure \ref{fig:f2} larger (smaller) values of \Lop\ 
are represented by  redder (bluer) colours. We can see that there is no clear suggestion 
that the scatter in driven by \Lop\ in any of the lines. Note however that the relatively 
narrow range in \Lop\ covered by our sample (from $\Lop=2.0\times10^{44}$ to $\Lop = 
1.6\times10^{46}$ ergs/s, corresponding to a factor of 80), together with the uncertainties
in our estimations of $f$, do not allow us to rule out this mechanism. 

Testing this model further, we found the combination of parameters $a_{1}$, $a_{2}$ and $f_{0}$
that best reproduce our \MAD\ measurements and the observed relation between $f$ and 
$\fwobs$ for the \Halpha\ line. To obtain dimensionless values for $a_{1}$ and $a_{2}$ we 
expressed $M_{\rm BH}$, $L_{\lambda}$ and $\fwhm$ in units of $10^{8}M_{\odot}$, 
$10^{44}\ergs$ and 1000 \kms, respectively. Taking $\alpha_{\rm line}=0.63$, as suggested by the 
observations (see Table \ref{tab:MAD}), we carried out a Monte-Carlo Markov Chain exploration of 
the parameter space of the model and found that $a_{1}=0.88$, $a_{2}=0.36$ and $f_{o}=0.51$ 
are able to reproduce our \MAD\ measurements with a scatter of 0.12 dex, preserving the 
experimental dependence of $R_{\rm BLR}$ on $L_{\lambda}$ as expressed in Equation \ref{eqn:rblr} 
with a scatter of 0.05 dex. At the same time the results are able to reproduce the observed 
$f$--\fwobs\ relation with a scatter of 0.11 dex (see Figure \ref{fig:E4}, which presents our
observations (black squares with error bars) together with the prescribed values for $f$ as given by 
Equation \ref{eqn:rad_factor} (coloured circles without error bars)). However, we also found that the residuals between 
the predicted values and the best fit to the correlation are heavily correlated with 
\Lop\ (r$_{\rm s}>0.63$, P$_{\rm s}<2\times10^{-5}$), as can be seen by the colour gradient of our
simulated points in the direction perpendicular to the correlation best fit in Figure \ref{fig:E4}. 
This bias is introduced by the explicit dependence of $f_{\rm rad}$ on $L_{\lambda}$ which is 
not observed in our sample, although notice that the error bars of our derived $f$ values are of the
order of, if not larger, than the expected dependence (see Figure \ref{fig:E4}). Finally, the 
dependency on \Lop\ vanishes when $\alpha_{\rm line}=0.5$. For this case, however, we were unable 
to reproduce any the observables. Extending our sample towards lower luminosities should yield 
the final test to be able to confidently conclude whether this model can be the driving 
mechanism for the observed $f$--\fwobs\ correlation.

Further constraints to the models evaluated here may be provided by the new method to determine \Mbh\ based on the spectropolarimetry of the broad lines\cite{Afanasiev2015}. This method takes advantage of the scattering in the dusty structure of the light coming from the BLR. The radial velocity gradient  in the BLR induces a gradient in the  position angle of polarization  across the broad emission line profiles. The advantage of this method is that it provides \Mbh\ estimations that are independent of the BLR inclination.  Therefore, the application of this method  will eventually allow us to differentiate between the effects of inclination and radiation pressure.

\end{document}